%% file: WAlg_HigherSpin.tex
\RequirePackage{ifpdf}
\documentclass[12pt]{article}

\input{gdefn.tex}

\begin{document}

\parindent=12pt

\begin{center}

{\Large \bf $W(0,b)$ algebra and the dual theory of 3D asymptotically flat higher spin gravity}
\end{center}

\baselineskip=18pt

\bigskip

\centerline{Nabamita Banerjee$^{a}$, Arindam Bhattacharjee$^{b}$, Surajit Biswas$^{a}$, Arpita Mitra$^{c}$,}
\centerline{and Debangshu Mukherjee$^{d}$}

\bigskip

\centerline{\it $^{a}$Indian Institute of Science Education and Research
Bhopal}
\centerline{\it Bhopal Bypass Road, Bhauri, Bhopal 420 066, Madhya Pradesh, India.}
\medskip
\centerline{\it $^{b}$Harish-Chandra Research Institute,}
\centerline{\it a CI of Homi Bhaba National Institute,}
\centerline{\it Chhatnag Road, Jhusi, Prayagraj 211019, Uttar Pradesh, India.}
\medskip
\centerline{\it $^{c}$Department  of  Physics,  Pohang  University  of  Science  and  Technology}
\centerline{\it  Pohang  37673,  Korea.}
\medskip
\centerline{\it $^{d}$Department of Physics, Indian Institute of Technology Kanpur}
\centerline{\it Kalyanpur, Kanpur 208016, India}

\bigskip

\centerline{\tt E-mail: nabamita@iiserb.ac.in, arindambhattacharjee@hri.res.in,}
\centerline{\tt surajit18@iiserb.ac.in, arpitamitra89@gmail.com, debangshu@iitk.ac.in}

\vskip .6cm
\medskip

\vspace*{4.0ex}

\centerline{\bf Abstract} \bigskip

\noindent 
BMS algebra in three spacetime dimensions can be deformed into a two parameter family of algebra known as $W(a,b)$ algebra. For $a=0$, we show that other than $W(0,-1)$, no other $W(0,b)$ algebra admits a non-degenerate bilinear and thus one can not have a Chern-Simons gauge theory formulation with them. However, they may appear in a three-dimensional gravity description, where we also need to have a spin 2 generator, that comes from the $(a=0,b=-1)$ sector. In the present work, we have demonstrated that the asymptotic symmetry algebra of a spin 3 gravity theory on flat spacetime has both the $W(0,-1)$ and $W(0,-2)$ algebras as subalgebras. We have also constructed a dual boundary field theory for this higher spin gravity theory by using the Chern-Simons/Wess-Zumino-Witten correspondence.

\vfill\eject

\tableofcontents

\section{Introduction}
BMS (Bondi-Metzner-Sachs) symmetry group describes the behavior of gravitational fields at null infinity, which is the asymptotic boundary for all massless fields in flat spacetimes \cite{BMvdB:1962, Sachs:1962, Barnich:2011mi}. In the framework of flat space holography, BMS symmetry plays a significant role. It is generated by killing vector fields which preserve the fall-off conditions of massless fields at null infinity. It encompasses both large gauge transformations, which involve supertranslations and superrotations. These symmetries are associated with an infinite number of conservation laws \cite{Strominger:2013jfa,He:2014laa} and physical observables at null infinity \cite{Saha:2019tub,Fernandes:2020tsq}. The recent formulation of celestial CFT with underlying BMS symmetry explores the dualities between four dimensional gravitational theories in flat spacetime and certain codimension two conformal field theories (CFTs) defined on the celestial sphere at null infinity \cite{Pasterski:2017kqt}. It provides a powerful framework to analyze the infrared structure of gravitational scattering amplitudes by studying the soft behavior of radiation \cite{Fotopoulos:2019vac, 
 Banerjee:2021uxe}. Similar aspects however are not well explored in the context of three space time dimensions. Since in three spacetime dimensions, any Fronsdal’s gauge field with ``spin" $s>1$ has no dynamical degrees of freedom, thus for a theory of gravity, we do not have bulk graviton and hence no scattering phenomenon is possible. However, there exist non-trivial boundary gravitons and one can construct the dual field theory of asymptotically flat gravity (or in the presence of a cosmological constant) utilizing a gauge theory formulation, namely the Chern-Simons formulation on a compact manifold \cite{Witten:1988hc}. In the  Chern-Simons formulation of gravity, by imposing suitable boundary conditions, one can further obtain a Wess-Zumino-Witten theory at the boundary from the bulk Chern-Simons theory with a (non) compact gauge group \cite{Witten:1988hf,Moore:1989yh,Elitzur:1989nr}. For pure gravity in asymptotically flat spacetime, one has to choose ISO(2,1) gauge group and impose suitable boundary conditions at null infinity to obtain the dual WZW model invariant under infinite dimensional BMS symmetries \cite{Barnich:2013yka,Banerjee:2015kcx}. This analysis has further been extended to three dimensional asymptotically flat gravity theories in the presence of (extended) supersymmetries in \cite{Barnich:2015sca, Banerjee:2016nio, Banerjee:2018hbl, Banerjee:2019lrv, Banerjee:2019epe,  Banerjee:2021uxl}.

 BMS algebra in three spacetime dimensions can be deformed into a family of algebra called $W(a,b)$ algebra with two arbitrary real parameters $a$ and $b$ \cite{Parsa:2018kys}. The deformation is introduced in the commutator between supertranslation and superrotation since it cannot be computed from the Hochschild-Serre factorization theorem known for deformations of finite dimensional algebras. The parameter $b$ in $W(a,b)$ algebra can be interpreted as a higher spin representation of Witt algebra and can be any real number. $a$ changes the periodicity of primary fields and ranges between $0$ and $1/2$. For different values of $a$ and $b$, we get different classes of algebras which may be associated with completely different asymptotic symmetries in two or three spacetime dimensions. In particular, $W(0,-1), W(0,0)$ and $W(0,1)$ algebras correspond to $\bmst$ algebra \cite{Barnich:2006av}, (centerless) Virasoro-Kac-Moody algebra \cite{Detournay:2012pc} and $\mathfrak{bms}_{2}$ algebra \cite{Afshar:2019axx}. Imposing different boundary conditions on the fields may give different asymptotic symmetries in a theory for an algebra with the same values of parameters. For instance, while $W(0,0)$ algebra can be obtained as an asymptotic symmetry algebra of AdS$_3$ with Comp\`{e}re-Song-Strominger boundary conditions \cite{Compere:2013bya}, it can also be realized as near horizon symmetry of BTZ black holes \cite{Afshar:2015wjm}.   $W(0,b)$ algebra can also be viewed as symmetries of Galilean field theories \cite{Chen:2019hbj, 10.1063/1.532067}. 
If one further deforms the $W(a,b)$ algebra and its central extension for generic values of the parameters, one gets back the same algebra with shifted parameters implying the rigidity of this algebra. 

Although some classes of $W(a,b)$ algebra are realized as physical symmetries, our first goal is to put constraints on the parameters $a$ and $b$ such that they admit a gauge theory formulation. For this purpose, we find which class of $W(a,b)$ algebra contains a closed global subalgebra and further admits a non-degenerate invariant bilinear or quadratic Casimir for arbitrary $a, b$,  thus allowing a Chern-Simons formulation with the corresponding gauge group. Interestingly, we ended up with a no-go result with only $W(0,-1)$ being the possible candidate. On the other hand, if one wants to obtain the three dimensional asymptotically flat Einstein gravity from the Chern-Simons theory it inevitably requires a spin 2 generator. It is well known that for $a=0, b=-1$, we have a spin 2 supertranslation generator and the asymptotic symmetries are BMS symmetries with appropriate boundary conditions imposed on the Chern-Simons gauge field. Thus one can possibly realize $W(0,b)$ algebra with integer values of $b$ as a subalgebra of higher spin extension of $W(0,-1)$ algebra. Here we consider the next simple possible integer value of $b=-2$, i.e. $W(0,-2)$ algebra and study its appearance in the asymptotic symmetries for spin 3 generalization of Einstein's gravity theory. 

Higher spin theory in four and higher dimensional curved spacetime was introduced by Vasiliev as a way to avoid all no-go theorems \cite{Vasiliev:1990en}. This theory is a higher spin extension of general relativity in which, in a constant curvature background, an infinite tower of massless higher spin fields are coupled together in a consistent manner. Higher spin theories in flat and AdS spacetime have connections to various areas of physics, including string theory, conformal field theory, and holography and are also expected to have important implications in cosmology \cite{Giombi:2009wh, Maldacena:2011jn, Campoleoni:2010zq, Vasiliev:2012vf, Bekaert:2022poo}. However, three space time dimensions are special, since the Weyl curvature vanishes for any gravitational background and is simpler in technical aspects. This suggests that in three space time dimensions, one can possibly avoid the no-go results for minimal coupling and can truncate the theory with a finite number of higher spin couplings \cite{Aragone:1983sz}. Moreover in three dimensions one can construct asymptotically flat higher spin gravity theories from an In\"on\"u-Wigner contraction of the
Vasiliev’s theory in AdS space \cite{Campoleoni:2021blr}. One does not have to worry about the no-go theorems and higher spin theories can be constructed just by generalizing the Chern-Simons formulation of gravity. 

Building on the above idea, in this paper we have constructed a dual field theory for spin 3 generalizations of pure Einstein's gravity in the bulk. The dual theory lies on the asymptotic null boundary of the bulk space time. We have utilized the equivalence between the theory of flat gravity, a Chern-Simons theory with ISO(2,1) gauge group and a chiral WZW model on the boundary of a three dimensional flat spacetime to write the dual theory. We have further gauged the WZW model and expressed it using the reduced phase space formalism. The resultant theory is a spin 3 generalization of a flat limit of the Liouville theory. To our best knowledge, this is the first ever result on the dual description of a three dimensional higher spin generalization of Einstein's gravity. 

The paper is organized as follows. In Section \ref{sec2} we analyze constraints on the class of $W(a,b)$ algebra which can admit a gauge theory formulation. In Section \ref{sec3} we revisit the Chern-Simons formulation of three dimensional gravity and construct the asymptotic charge algebra for flat space spin 3 gravity. We find that the asymptotic charge algebra contains $W(0,-2)$ as a sub-algebra. In addition, we also verify the charge algebra with that of the  In\"on\"u-Wigner contraction of two copies of AdS$_3$ spin 3 asymptotic charge algebras, presented in \cite{Campoleoni:2021blr} and find a perfect agreement. Finally, we compare our results with that in the existing literature \cite{Afshar:2013vka}. Section \ref{sec5} contains the main result of the paper, where we have presented the construction of the dual theory of spin 3 extensions of Einstein's gravity. The resultant theory is a simple generalization of the flat limit of Liouville theory. We conclude with a discussion and outlook in Section \ref{sec6}. The appendices contain relevant computational details.
\section{\texorpdfstring{$W(a,b)$}{} algebra, its global subalgebra and implications in 3D flat gravity}\label{sec2}

Previous works \cite{Safari:2019zmc, Banerjee:2022abf} have studied deformations of the $\bmst$ algebra, dubbed as $W(a,b)$ algebra where the deformation is encoded in the parameters $a$ and $b$. The $W(a,b)$ algebra can be written out explicitly as follows
\begin{equation} 
\label{eq:W(a,b)algebra}
\begin{split}
 & [{J}_{m},{J}_{n}]=(m-n){J}_{m+n}\ ,\\
 &[{J}_{m},{P}_{n}]=-(n+bm+a){P}_{m+n}\ ,\\
 &[{P}_{m},{P}_{n}]=0\ ,
\end{split}
\end{equation}
where $J_n$ and $P_n$ physically correspond to superrotation and supertranslation generators respectively. The index of superrotations $J_m$ can take only integer values while that of $P_m$ can be integer or half-integer. The first relation above in \eqref{eq:W(a,b)algebra} is the centerless Virasoro algebra which admits a closed subalgebra consisting of $J_{\pm 1}, J_0$. Further, if \eqref{eq:W(a,b)algebra} admits closure of finitely many supertranslation generators $P_n, ~{n= N_1,...,N_k, N_1<N_k}$, then those along with $J_{\pm 1}, J_0$ will form a global subalgebra. 
To explore the possibility, we focus on the commutator $[J_m, P_n]$ for $m = \pm 1, 0$. 
\begin{itemize}
    \item For $m=0$, we get, $[J_0, P_n]=-(n+a)P_n$. Thus, the action of $J_0$ on the supertranslation generator keeps the index invariant. 
    \item For $m=1$, we get, $[J_1, P_n]=-(n+b+a)P_{n+1}$. Thus at the level of the algebra, $J_1$ essentially increases the index of the supertranslation by one. Demanding the existence of a global subalgebra implies, we must have $[J_1, P_{N_k}]=0$ which leads us to $N_k =-a-b$.
    \item For $m=-1$, we get, $[J_{-1}, P_n]=-(n-b+a)P_{n-1}$ demonstrating that $J_{-1}$ reduces index of $P_n$ by one. Thus, for $n=N_1$, we must have, $N_1-b+a=0$ implying $N_1=-a+b$.
\end{itemize}
Since by assumption, $N_1< N_k$, we must have $b<0$ and thus the so-called global sub-sector will consist of $2(|b|+2)$ elements---$2|b|+1$ supertranslation labelled as $P_{-a+b},..,P_{-a-b}$ and 3 superrotations $J_{-1},J_0$ and $J_{+1}$. Thus, we see that existence of a global subalgebra of $W(a,b)$ algebra is subject to the deformation parameters $a$ and $b$ being integer or half-integer while $b$ must also be strictly negative. As discussed in the introduction, presently  we are interested in the particular $W(0,-2)$ deformation. The algebra is given as ,
\begin{equation} 
\label{eq:W(0,-2)algebra}
\begin{split}
 & [{L}_{\alpha},{L}_{\beta}]=(\alpha-\beta){L}_{\alpha+\beta}\ ,\\
 &[{L}_{\alpha},{\mathcal{V}}_{q}]=(2\alpha-q){\mathcal{V}}_{\alpha+q}\ ,\\
 &[{\mathcal{V}}_{p},{\mathcal{V}}_{q}]=0\ ,
\end{split} .
\end{equation}
This is an infinite dimensional algebra. The global sector consisting of $L_{\pm 1}, L_0, {\mathcal{V}}_{\pm 2}, {\mathcal{V}}_{\pm 1}, {\mathcal{V}}_0$ forms a close subalgebra. In the next section, we will explore the importance of this algebra in the context of three dimensional asymptotically flat theories. Our root to understand this aspect would be a three dimensional Chern-Simons formulation of gravity theories.

\subsection{Invariant bilinear for global sector of \texorpdfstring{$W(a,b)$}{} algebra : A No-Go Theorem}
The global sector of the $W(a,b)$ algebra defined in \eqref{eq:W(a,b)algebra} can be realized as a symmetry group of a Chern-Simons gauge theory provided it admits a non-degenerate quadratic Casimir for the same. The presence of the non-degenerate quadratic Casimir in turn dictates the existence of so-called symmetric supertrace elements of the theory. These supertrace elements are required while we express the standard Chern-Simons action on the gauge manifold, by expanding the Chern-Simons gauge field in terms of component fields along the gauge group generators.

The most general ansatz for a possibly non-degenerate Casimir can be written as,
\begin{align}
\label{eq:casimiransatz}
    \cC^2 &=  \sum_{r=-1}^{+1} \sum_{s=-1}^{+1} \alpha^{rs} J_r J_s +  \sum_{M=-a+b}^{-a-b} \sum_{N=-a+b}^{-a-b} \beta^{MN} P_{M} P_{N} + \sum_{M=-a+b}^{-a-b} \sum_{r=-1}^{+1} \gamma^{Mr} P_{M} J_r
\end{align}
where $\alpha$ and $\beta$ are $3 \times 3$ and $(2|b|+1) \times (2|b|+1)$ square matrices while $\gamma$ is $(2|b|+1) \times 3$ 
matrix. Note that $\gamma$ is square matrix for $\bmst$ case ($a=0, b=-1$). From now on, for all our subsequent calculations, we will use small Latin indices i.e. $r,s,p,q$ to denote the labels of the global sector of superrotations composed of $\{J_{\pm 1},J_0\}$ and capital Latin indices $K,L,M,N$ to denote the $2|b|+1$ a number of global generalized supertranslation generators $P_M$.

Alternatively, one could have also written the Casimir as
\begin{equation}
\label{eq:CasimirMatrix}
    \cC^2=\begin{pmatrix}
    P_M & J_r
    \end{pmatrix}
    {\begin{pmatrix}
    \beta^{M,N} & \gamma^{M,s}\\\gamma^{r,N} & \alpha^{rs}
    \end{pmatrix}}
    \begin{pmatrix}
    P_N\\J_s
    \end{pmatrix} \equiv
    \begin{pmatrix}
    P_M & J_r
    \end{pmatrix}
    \Big[\cG^{ab}\Big]
    \begin{pmatrix}
    P_N\\J_s
    \end{pmatrix}\ ,
\end{equation}
where the metric in the field space $\cG^{ab}$ must be non-degenerate (non-vanishing determinant) in order for the gravity theory to admit a Chern-Simons description. Our goal is to determine the most generic form of the matrices $\alpha, \beta$ and $\gamma$ such that the matrix $\cG^{ab}$ has a non-zero determinant. 

By definition, since Casimir commutes with all generators, we must have
\begin{equation}\label{casimir:cons}
    [\cC^2, J_p]=[\cC^2, P_L]=0 \ \forall\ p \in \{-1,0,+1\}\ \text{and}\ L \in \{-a+b,..,-a-b\}\ . 
\end{equation}
These relations boils down to the following, 
\begin{equation}
\label{eq:CasimirJpcommutator}
\begin{aligned}
&\sum_{r=-1}^{+1} \sum_{s=-1}^{+1} \alpha^{rs}[(s-p)J_rJ_{s+p}+(r-p)J_{r+p}J_s] \\
&\hspace{0.25cm}+ \sum_{M=-a+b}^{-a-b} \sum_{N=-a+b}^{-a-b} \beta^{MN}[(N+bp+a)P_M P_{N+p} +(M+bp +a)P_{M+p}P_N ] \\
&\hspace{0.35cm}+\sum_{M=-a+b}^{-a-b} \sum_{r=-1}^{+1} \gamma^{Mr}[(r-p)P_M J_{r+p} + (M+bp+a)P_{M+p}J_r]= 0\ , 
\end{aligned}
\end{equation}
and
\begin{equation}
\label{eq:CasimirPLcommutator}
\begin{aligned}
&\sum_{r=-1}^{+1} \sum_{s=-1}^{+1}\alpha^{rs}\{(L+br+a)P_{r+L}J_s+(L+bs+a)J_r P_{s+L}\}\\
&\hspace*{5.5cm}+\sum_{M=-a+b}^{-a-b} \sum_{r=-1}^{+1}\gamma^{Mr}(L+br+a)P_MP_{r+L}=0\ .
\end{aligned}
\end{equation}
Eq. \eqref{eq:CasimirJpcommutator} dictates that $\alpha$ and $\beta$ satisfy the following conditions,
\begin{equation}\label{alpha:condition}
    \sum_{r,s=-1}^{+1}(\alpha^{rs}+\alpha^{sr})(s-p)J_rJ_{s+p}+\sum_{r,s=-1}^{+1}\alpha^{sr}(s-p)(s-r+p)J_{r+s+p}=0
\end{equation}
\begin{equation}\label{beta:condition}
    \begin{aligned}
       \sum_{M,N=-a+b}^{-a-b}(\beta^{MN}+\beta^{NM})(N+bp+a)P_MP_{N+p}=0
    \end{aligned}
\end{equation}
Since $\cG_{ab}$ is symmetric, hence $\alpha$ will always be vanishing to satisfy \eqref{alpha:condition} while non-zero components of $\beta$ can exist for specific values of $a$ and $b$. Considering the case of $a=0$, \eqref{beta:condition} for $p=0,\pm 1$ implies that some components of $\beta$\footnote{particularly $\beta^{-n,n},\cdots,\beta^{-11}, \beta^{00}$, where $n$ is any integer, are non zero.} remains non-vanishing for integer values of $b$. Previously we have realized that for a generic $W(a,b)$ algebra, $b$ has to be negative for finding a well defined global subalgebra. Hence for finding a non-trivial $\beta$ matrix, we are restricted to only the choice of $b$ being a negative integer with vanishing $a$. But since $\alpha$ is zero, having a non-vanishing $\beta$ does not imply the existence of a non-degenerate bilinear for the corresponding algebra unless we have a non-vanishing $\gamma$ matrix. In fact,
since components of $\alpha$ are zero for any $W(a,b)$ algebra, $\cG_{ab}$ can still be non singular with (non)vanishing $\beta$ if some symmetric components of $\gamma$ exist. In this context, one can find the non-vanishing $\gamma$ components from \eqref{eq:CasimirPLcommutator} if the following relation holds :
\begin{equation}\label{eq:gamma}
    \sum_{M=-a+b}^{-a-b} \sum_{r=-1}^{+1}\gamma^{Mr}(L+br+a)P_MP_{r+L}=0 .
\end{equation}

For $a=0$, from \eqref{eq:gamma} with $L$ taking values between $-b$ to $b$ one finds that $\gamma$ only exist for $b= - 1$. Hence the absence of non-degenerate $\cG^{ab}$ for other values of $b$ poses obstruction in a gauge theory formulation with the gauge symmetry algebra being the global $W(a,b)$ where $b\neq - 1$. Thus we have established the following No-Go theorem : 

{\it Other than $W(0,-1)$, no other $W(0,b)$ algebras admits a closed global subalgebra with non-degenerate invariant bilinears and hence can not be used as a gauge group of a three dimensional Chern-Simons theory.}

However one can realize $W(0,b)$ algebras with arbitrary negative integer valued $b$ even other than $-1$ as a sub-algebra of the bigger symmetry group of a gauge theory. We discuss this possibility in the following sections.

\section{Implications of \texorpdfstring{$W(a,b)$}{} in 3D asymptotically flat gravity}\label{sec3}
In three dimensional asymptotically flat spacetime one always require a spin 2 generator along with the generators of Witt algebra to obtain Poincar\'e symmetry in the bulk. Hence one can realize any $W(0,b)$ algebra with $b<-1$ as a higher spin generalization of $W(0,-1)$ or equivalently the $\bmst$ algebra. Next, we will construct the Chern-Simons action for spin 3 generalizations of $iso(2,1)$ algebra following a basis change.
\subsection{Chern-Simons formulation of three dimensional asymptotically flat spin 3 gravity}
\label{sec:CSformulation}
In the Chern-Simons formulation of gravity, a $2+1$ dimensional (super)gravity theory can be expressed in terms of Chern-Simons action as:
\begin{equation}\label{csaction}
I_{CS}[{\cal A}] = \frac{k}{4\pi}\int_M \langle \mathcal{A} , d\mathcal{A} +\frac{2}{3} \mathcal{A}^2\rangle .
\end{equation}
 with an appropriate gauge symmetry group, suitable for (super) gravity. Here the gauge field ${\cal A}$ is a Lie-algebra-valued one form and $\langle,\rangle$ represents metric
in the field space that one obtains from the supertrace elements on
the Lie algebra space. $M$ is a three dimensional manifold and $k$ is the level for the Chern-Simons theory. For a theory of gravity, it takes the value $k=\frac{1}{4G}$ where $G$ is Newton's constant.  We express $\mathcal{A} = \mathcal{A}^a_\mu T_a dx^ \mu$ where ${T_a}$ are
a particular basis of the Lie-algebra associated with bulk symmetries. The equation of motion for the gauge field $A$ is given as: $F\equiv d\mathcal{A} + \mathcal{A} \wedge \mathcal{A} =0.$  

Conventionally, it is easiest to think of flat space-times as the \emph{infinite radius} limit of $AdS$ spacetimes. In spirit, this is precisely what the \.{I}n\"{o}n\"{u}-Wigner contraction achieves, in the context of the corresponding symmetry algebras. Asymptotic symmetry algebra of three-dimensional $AdS$ gravity with spin 3 fields, (essentially given by two independent copies of the $sl(3,\mathbb{C})$ i.e. $sl(3,\mathbb{C}) \bigotimes sl(3,\mathbb{C})$) has been studied earlier in \cite{Campoleoni:2010zq} where each copy of $sl(3,\mathbb{C})$ is written as
\begin{align}
    &[\mathcal{L}_{\alpha},\mathcal{L}_{\beta}]= (\alpha-\beta)\mathcal{L}_{\alpha+\beta}\ , \nonumber\\
    &[\mathcal{L}_{\alpha},{W}_{q}]= (2\alpha-q){W}_{\alpha+q}\ , \nonumber\\
    \label{eq:sl3Cdc}
    &[{W}_{p}, {W}_{q}]= \frac{\sigma}{3}[(p-q)(2p^{2}+2q^{2}-pq-8)\mathcal{L}_{p+q}\ .
\end{align}
Here the Greek indices associated with the $sl(2,\mathbb{C})$ generators $(\alpha, \beta) \in \{-1,0,+1\}$ while those associated with the spin-3 fields are denoted by the Latin indices $(p,q) \in \{-2,-1,..,+2\}$ and $\sigma$ is a non zero parameter, that denotes the higher spin coupling. Performing a linear transformation as
\begin{eqnarray}
\label{eq:IWb1-1}
L_\alpha &=&\mathcal{L}_\alpha -\bar{\mathcal{L}}_{-\alpha}\ ,\\
M_\alpha &=& \frac{1}{l}(\mathcal{L}_\alpha +\bar{\mathcal{L}}_{-\alpha})\ ,\\
U_p&=&W_p-\bar{W}_{-p}\ ,\\
\label{eq:IWb1-4}
V_p&=&\frac{1}{l}(W_p+\bar{W}_{-p})\ .
\end{eqnarray}
(the generators $\bar{\mathcal{L}}_{\alpha}$ and $\bar{W}_p$ are the ones associated to the second copy of $sl(3,\mathbb{C})$) and subsequently taking the $l \rightarrow \infty$ limit leads us to
\begin{align}\label{HSBA}
    [{L}_{\alpha},{L}_{\beta}]=&(\alpha-\beta){L}_{\alpha+\beta}\ , \nonumber \\
    [{L}_{\alpha},{M}_{\beta}]=&(\alpha-\beta){M}_{\alpha+\beta}\ , \nonumber \\
[{L}_{\alpha},U_{q}]=&(2\alpha-q)U_{\alpha+q}\ , \nonumber \\
[{L}_{\alpha},V_{q}]=&(2\alpha-q)V_{\alpha+q}\ , \nonumber \\
[{M}_{\alpha},U_{q}]=& (2\alpha-q)V_{\alpha+q}\ ,
   \nonumber \\
[U_{p},U_{q}]=&\frac{\sigma}{3}(p-q)(2p^{2}+2q^{2}-pq-8)L_{p+q}\ ,\nonumber \\
[U_{p},V_{q}]=&\frac{\sigma}{3}(p-q)(2p^{2}+2q^{2}-pq-8){M}_{p+q}\ ,
\end{align}
which is the form used in the analysis of asymptotically flat higher spin gravity theories \cite{Afshar:2013vka}. We will call this algebra $\mathfrak{fhs}(3)$ (with $\mathfrak{fhs}$ denoting \emph{flat-space higher spin}). Here, $L_{\alpha}$ and $V_p$ form a subalgebra which is exactly the same as the global sector of $W(0,-2)$ algebra discussed in the previous section. For the sake of brevity, we will be calling the above basis as the $sl(2,\mathbb{R})$ embedding of $\mathfrak{fhs}(3)$ or $\mathfrak{fhs}(3)_{sl(2)}$ in short. The nomenclature of this basis is obvious from the first relation in \eqref{HSBA} which is precisely the $sl(2,\mathbb{R})$ algebra. Interestingly, the above algebra \eqref{HSBA} has a non-degenerate invariant bilinear given as \cite{Afshar:2013vka}, 
\begin{equation}\label{BL}
    \langle L_{\alpha} ,M_{\beta} \rangle= -\frac{1}{2}\rho_{\alpha \beta}, \quad  \langle U_{p},V_{q}\rangle = -\frac{\sigma}{2} \kappa_{pq},
\end{equation} 
where $\rho_{\alpha\beta}= \text{anti-diagonal} (1, -\frac{1}{2},1)$ and $\kappa_{pq}= \text{anti-diagonal} (-4, 1,-\frac{2}{3},1,-4)$. Further, this is the most generic possible non-degenerate invariant bilinear possible for the given algebra.

Although an asymptotic symmetry analysis was done in \cite{Afshar:2013vka} by imposing appropriate boundary conditions in terms of an equivalent Chern-Simons theory describing the higher spin gravity theory, the final gravity action seems to take a complicated form when written in the first-order formalism. There is in fact a different basis for $\mathfrak{fhs}(3)$ which can be written by embedding the $so(2,1)$ algebra i.e. (algebra of Lorentz group in 3 spacetime dimensions). The advantage of working on this basis is that one can manifestly see the conventional \emph{first-order gravity term} $e^a \wedge R_a$ on writing the Chern-Simons action appropriately in terms of vielbeins and spin connections. Explicitly, this basis of $\mathfrak{fhs}(3)$ i.e. $\mathfrak{fhs}(3)_{so(1,2)}$ can be written as 
\begin{align}
\label{eq:JPJabPabbasis}
    &[J_{a},J_{b}]=\epsilon_{abc}J^{c},~[J_{a},P_{b}]=\epsilon_{abc}P^{c},~[P_{a},P_{b}]=0\ , \nonumber\\
    &[J_{a},J_{bc}]=\epsilon^m_{a(b}J_{c)m},~[J_{a},P_{bc}]=\epsilon^m_{a(b}P_{c)m},~[P_{a},J_{bc}]=\epsilon^m_{a(b}P_{c)m},~[P_{a},P_{bc}]=0\ ,\nonumber\\
    &[J_{ab},J_{cd}]=\sigma(\eta_{a(c}\epsilon_{d)bm}+\eta_{b(c}\epsilon_{d)am})J^{m},~[J_{ab},P_{cd}]=\sigma(\eta_{a(c}\epsilon_{d)bm}+\eta_{b(c}\epsilon_{d)am})P^{m}, [P_{ab},P_{cd}]=0\ ,
\end{align}
where the indices $(a,b,c) \in \{0,1,2\}$. Again, the first relation in the above basis makes the $so(2,1)$ sub-algebra manifest. The two-index operators $J_{ab}$ and $P_{ab}$ correspond to higher spin generators. In the above, $\epsilon_{012}=-\epsilon^{012}=1$ while the indices $a$ and $b$ are raised/lowered with respect to the metric $\eta_{ab}=\text{diag}(-1,+1,+1)$. Both operators $J_{ab}$ and $P_{ab}$ are symmetric in its indices and follows the \emph{tracelessness conditions} i.e.
\begin{equation}
    J_{00}=J_{11}+J_{22} \quad \text{and}\quad P_{00}=P_{11}+P_{22}\ . 
\end{equation}
In this basis, one can express the Chern-Simons gauge field as,
 \begin{align}
 \label{eq:CSAeomegamap}
{\cal{A}}=E^{a}P_{a}+\Omega^{a}J_{a}+E^{ab}P_{ab}+\Omega^{ab}J_{ab}\ .
 \end{align}
Here $(E^a, \Omega^a)$ denotes the fields of the usual spin two sector whereas $(E^{ab}, \Omega^{ab})$ denotes the fields of the spin 3 sector. The supertrace elements of the generators are given by
 \begin{align}
     \langle P_{a},J_{b}\rangle=\eta_{ab}\ ,\ \langle P_{ab},J_{cd}\rangle=-2\sigma\eta_{ab}\eta_{cd}\ .
 \end{align}

The Chern-Simons action can be written in terms of the so-called vielbeins $E^a$, spin connections $\Omega^a$ and higher-spin fields $E^{ab}, \Omega^{ab}$ as,
\begin{align}
    I_{CS}=&\frac{k}{4\pi}\int_M \langle \mathcal{A} , d\mathcal{A} +\frac{2}{3} \mathcal{A}^2\rangle\\
    =&\frac{k}{4\pi}\int_{\mathcal{M}}\bigg[2E^{a}\Big(d\Omega_{a}+\frac{1}{2}\epsilon_{abc}\Omega^{b}\Omega^{c}-2\sigma\epsilon_{abc}\Omega^{bd}\Omega^{c}_{~d}\Big)\bigg.
    \\ &\hspace*{4.5cm}\bigg.-4\sigma E^{ab}\Big(d\Omega_{ab}+\epsilon_{cda} \Omega^{c}\Omega_{b}^{~d}+\epsilon_{cdb} \Omega^{c}\Omega_{a}^{~d}\Big)\bigg]\ .
\end{align}
The above action describes the coupling of spin 3 field that of gravity in asymptotically flat three space time dimensions \cite{Campoleoni:2010zq}. Here $\sigma$ denotes the coupling constant of the higher spin interactions. The equations of motion following from the above action are given by,
\begin{align}
&dE^{a}+\epsilon^{abc}\Omega_{b}E_{c}-4\sigma\epsilon^{abc}E_{bd}\Omega_{c}^{~d}=0\ ,\nonumber\\
    &d\Omega^{a}+\frac{1}{2}\epsilon^{abc}\Omega^{b}\Omega^{c}-2\sigma\epsilon^{abc}\Omega_{bd}\Omega_{c}^{~d}=0\ ,\nonumber\\
&dE^{ab}+\epsilon^{cda}\Omega_{c}E_{d}^{~b}+\epsilon^{cdb}\Omega_{c}E_{d}^{~a}+\epsilon^{cda}e_{c}\Omega_{d}^{~b}+\epsilon^{cdb}E_{c}\Omega_{d}^{~a}=0\ ,\nonumber\\
&d\Omega^{ab}+\epsilon^{cda}\Omega_{c}\Omega_{d}^{~b}+\epsilon^{cdb}\Omega_{c}\Omega_{d}^{~a}=0\ .
\end{align}

The two basis \eqref{HSBA} and \eqref{eq:JPJabPabbasis} are related by
\begin{equation}
    L_{\alpha}={\cal U}_{\alpha}^{\ a}J_a \quad \text{and} \quad M_{\alpha}={\cal U}_{\alpha}^{\ a}P_a
\end{equation}
where
\begin{equation}
    {\cal U}_{\alpha}^{\ a} = 
    \begin{pmatrix}
        -1 &-1&0\\
        0&0&1\\
        -1&1&0
    \end{pmatrix}\ .
\end{equation}
In the above $\alpha \in \{-1,0,+1\}$ label the row indices while $a \in \{0,1,2\}$ label the column indices. The higher spin generators $U_p$ and $V_p$ also have a nice structure since the $U_p$ generators mix the $J_{ab}$ generators while the $V_p$ generators mix exclusively between the $P_{ab}$ generators. The linear transformation connecting these two sets of generators can be explicitly written as
\begin{equation}
    \begin{pmatrix}
        U_{-2}\\
        U_{-1}\\
        U_0\\
        U_{+1}\\
        U_{+2}
    \end{pmatrix}=
    \begin{pmatrix}
        2&0&2&0&1\\
        0&-1&0&-1&0\\
        0&0&0&0&1\\
        0&-1&0&1&0\\
        -2&0&2&0&1
    \end{pmatrix}
    \begin{pmatrix}
        J_{01}\\
        J_{02}\\
        J_{11}\\
        J_{12}\\
        J_{22}
    \end{pmatrix}\ \
    \text{and}\ \
    \begin{pmatrix}
        V_{-2}\\
        V_{-1}\\
        V_0\\
        V_{+1}\\
        V_{+2}
    \end{pmatrix}=
    \begin{pmatrix}
        2&0&2&0&1\\
        0&-1&0&-1&0\\
        0&0&0&0&1\\
        0&-1&0&1&0\\
        -2&0&2&0&1
    \end{pmatrix}
    \begin{pmatrix}
        P_{01}\\
        P_{02}\\
        P_{11}\\
        P_{12}\\
        P_{22}
    \end{pmatrix}\ .
\end{equation}

In the following, we will revisit the construction of charge algebra corresponding to the asymptotic symmetries associated with spin 3 gravity by imposing suitable fall-off conditions on the Chern-Simons gauge field. A similar analysis was carried on in \cite{Afshar:2013vka}. We shall comment on the results obtained towards the end of this section.

\subsection{Construction of asymptotic algebra}
The presence of boundary in a theory always enhances the boundary symmetries. These enhanced symmetries at the boundary are in general known as the asymptotic symmetries of the corresponding theory. For three dimensional asymptotically flat Einstein (super) gravity, one can define a set of boundary conditions such that gives us (super) BMS as the asymptotic symmetry group. In the Chern-Simons formulation of (super)gravity in the ($u, r, \phi$) coordinates first we do a partial gauge fixing by imposing $\partial_{\phi}\mathcal{A}_r=0$ and then impose the boundary condition on
the gauge fields as,
\begin{align}
    \mathcal{A} = b^{-1} (a + d) b, ~~b=e^{\frac{r}{2} M_{-1}}
\end{align}
Here the choice of $b$ helps to factorize the $r$ dependence and  $a(u,\phi)=a_{\phi}(u,\phi) d\phi + a_{u}(u,\phi) du$ is the boundary gauge field implying the residual gauge freedom at the boundary. Consideration of the asymptotic flat metric,
\begin{equation}
    ds^2=\mathcal{M} du^2-2 dudr+\mathcal{N} du d\phi+r^2 d\phi^2\label{ASFM}
\end{equation}
dictates the boundary conditions fixing the components of $a$ accordingly. Thus for a spin 3 extended Einstein's gravity, we begin with,
\begin{align}\label{BC}
    a_{u} &= M_{1}-\frac{1}{4} \mathcal{M}(u,\phi)~M_{-1}+\mathcal{Q}(u,\phi) V_{-2} \notag\\
    a_{\phi} &= L_{1}-\frac{1}{4} \mathcal{M}(u,\phi) L_{-1} -\frac{1}{4} \mathcal{N}(u,\phi) M_{-1}+\mathcal{Q}(u,\phi)~ U_{-2} +\mathcal{O}(u,\phi)~ V_{-2}
    \end{align}
where $\mathcal{Q}, \mathcal{O}$ are fields associated with the higher spin generators. Since they do not generate any spacetime symmetries these fields are not involved in the metric \eqref{ASFM}. All fields involved in \eqref{BC} have to satisfy the gauge equation of motion hence implying constraints on these fields.  

The asymptotic symmetries correspond to the set of gauge transformations that preserves the equation of motion and variation of the boundary gauge field. The equation of motion and the gauge variation equation, respectively are given by,
\begin{align}
\label{eq:EOMGaugeV}
    da+\frac{1}{2}[a,a]=0, ~~\delta a=d\Lambda+[a,\Lambda]\ ,
\end{align}
where $\Lambda$ is the most general gauge parameter,
\begin{align}\label{GP}
\Lambda=\Upsilon^{n}L_{n}+\xi^{n}M_{n}+\epsilon^{p} U_{p}+\eta^{p} V_{p}
\end{align}
and $\Upsilon^{n}, \xi^{n}, \epsilon^{p}$ and $\eta^{p}$ are arbitrary fields depending on both $u$ and $\phi$ where the index $n \in \{-1,0,1\}$ and index $p \in \{-2,-1,0,1,2\}$.

The equations of motion explicitly in terms of the fields $\mathcal{M}$, $\mathcal{N}$, $\mathcal{O}$ and $\mathcal{Q}$ takes the form,
\begin{align}
\label{eq:CSEOM}
\partial_{u}\mathcal{M}=\partial_{u} \mathcal{Q}=0\ ,\quad ~~\partial_{u}\mathcal{O}=\partial_{\phi}\mathcal{Q}\ , \quad ~~\partial_{u}\mathcal{N}=\partial_{\phi}\mathcal{M}\ .
\end{align}
The above immediately suggests that $\mathcal{M}$ and $\mathcal{Q}$ are purely functions of $\phi$. Choosing $\mathcal{M}$ and $\mathcal{Q}$ to be independent functions, we can express the other two functions, namely $\mathcal{N}$ and $\mathcal{O}$ in terms of the independent functions thus fixing the explicit $u$ dependence of the asymptotic metric \eqref{ASFM}. To be more explicit, we can write,
\begin{align} \mathcal{N}(u,\phi)=\mathcal{L}(\phi)+u\partial_{\phi}\mathcal{M}(\phi)\ ,\quad \mathcal{O}(u,\phi)=\mathcal{U}(\phi)+u\partial_{\phi}\mathcal{Q}(\phi)\ ,
\end{align}
where $\mathcal{L}(\phi)$ and $\mathcal{U}(\phi)$ are arbitrary smooth functions of $\phi$.

We now turn our attention to the gauge variation equation (second equation of \eqref{eq:EOMGaugeV}). As one would expect, this gives us a large set of equations relating \emph{most} of the gauge field parameters--in fact, as we will see, we are precisely left with four independent gauge parameters, namely $\Upsilon^1, \xi^1, \epsilon^2$ and $\eta^2$. The variation of the field component $a_u$ leads us to
\begin{equation}
    \partial_{u} \Upsilon^{n}=0\ ,\quad \partial_{u}\epsilon^{p}=0\ ,
\end{equation}
along with other equations relating to various gauge functions. The above immediately implies both the arbitrary gauge functions $\Upsilon^n$ and $\epsilon^p$ must be functions of $\phi$ exclusively. 

The variation of $a_{\phi}$ further leads to another set of equations relating various gauge functions. As claimed before, considering $\Upsilon^1, \xi^1, \epsilon^2$ and $\eta^2$ to be the independent fields, the other gauge parameters can be expressed as
\begin{eqnarray}
    \Upsilon^{0}&=&-\partial_{\phi}\Upsilon^{1}\ ,\nonumber\\
    \Upsilon^{-1}&=&\frac{1}{2}\partial^{2}_{\phi} \Upsilon^{1}-\frac{1}{4}\mathcal{M}\Upsilon^{1}+8\sigma\mathcal{Q}\epsilon^{2}\ ,\nonumber\\
    \xi^{0}&=&-\partial_{\phi}\xi^{1}\ ,\nonumber\\
    \xi^{-1}&=&\frac{1}{2}\partial^{2}_{\phi} \xi^{1}-\frac{1}{4}\mathcal{M}\xi^{1}-\frac{1}{4}\mathcal{N}\Upsilon^{1}+8\sigma\mathcal{Q}\eta^{2}+8\sigma\mathcal{O}\epsilon^{2}\ ,\nonumber\\
    \epsilon^{1}&=&-\partial_{\phi}\epsilon^{2}\ ,\nonumber\\
    \label{eq:gaugedependent}
    \epsilon^{0}&=&\frac{1}{2}\partial^{2}_{\phi} \epsilon^{2}-\frac{1}{2}\mathcal{M}\epsilon^{2}\ ,\\
    \epsilon^{-1}&=&-\frac{1}{6}\partial_{\phi}^{3}\epsilon^{2}+\frac{1}{6}\partial_{\phi}\mathcal{M}\epsilon^{2}+\frac{5}{12}\mathcal{M}\partial_{\phi}\epsilon^{2}\ ,\nonumber\\
    \epsilon^{-2}&=& \frac{1}{24}\partial^{4}_{\phi}\epsilon^{2}-\frac{1}{24}\partial_{\phi}^{2}\mathcal{M}\epsilon^{2}-\frac{7}{48}\partial_{\phi}\mathcal{M}\partial_{\phi}\epsilon^{2}-\frac{1}{6}\mathcal{M}\partial_{\phi}^{2}\epsilon^{2}+\frac{1}{16}\mathcal{M}^{2}\epsilon^{2}+\mathcal{Q}\Upsilon^{1}\ ,\nonumber\\
    \eta^{1}&=&-\partial_{\phi}\eta^{2}\ ,\nonumber\\
    \eta^{0}&=&\frac{1}{2}\partial^{2}_{\phi} \eta^{2}-\frac{1}{2}\mathcal{M}\eta^{2}-\frac{1}{2}\mathcal{N}\epsilon^{2}\ ,\nonumber\\
    \eta^{-1}&=&-\frac{1}{6}\partial_{\phi}^{3}\eta^{2}+\frac{1}{6}\partial_{\phi}\mathcal{M}\eta^{2}+\frac{1}{6}\partial_{\phi}\mathcal{N}\epsilon^{2}+\frac{5}{12}\mathcal{M}\partial_{\phi}\eta^{2}+\frac{5}{12}\mathcal{N}\partial_{\phi}\epsilon^{2}\ , \nonumber\\
    \eta^{-2}&=& \frac{1}{24}\partial^{4}_{\phi}\eta^{2}-\frac{1}{24}\partial_{\phi}^{2}\mathcal{M}{\eta^2}-\frac{1}{24}\partial_{\phi}^{2}\mathcal{N}\epsilon^{2}-\frac{7}{48}\partial_{\phi}\mathcal{M}\partial_{\phi}\eta^{2}-\frac{7}{48}\partial_{\phi}\mathcal{N}\partial_{\phi}\epsilon^{2}-\frac{1}{6}\mathcal{M}\partial_{\phi}^{2}\eta^{2}\nonumber \\
      &&\hspace*{3.5cm} -\frac{1}{6}\mathcal{N}\partial_{\phi}^{2}\epsilon^{2}+\frac{1}{16}\mathcal{M}^{2}\eta^{2}+\frac{1}{16}\mathcal{M}\mathcal{N}\epsilon^{2}+\frac{1}{16}\mathcal{N}\mathcal{M}\epsilon^{2}+\mathcal{Q}\xi^{1}+\mathcal{O}\Upsilon^{1}\ .\nonumber
\end{eqnarray}
Furthermore, the above equations (obtained from gauge variations of $a_{\phi}$) can be used along with the equations obtained from the $a_u$ variation to obtain
\begin{equation}
    \partial_{u}\xi^{n}=\partial_{\phi}\Upsilon^{n}\ , \quad \partial_{u}\eta^{p}=\partial_{\phi}\epsilon^{p}\ .
\end{equation}
We have already seen that out of the four independent gauge parameters, $\Upsilon^1, \xi^1, \epsilon^2$ and $\eta^2$, $\Upsilon^1$ and $\epsilon^2$ are exclusively dependant on $\phi$. The above equation fixes the $u$-dependence of the other two functions $\xi^1$ and $\eta^2$ as follows 
\begin{align} \xi^{1}(u,\phi)=T(\phi)+u\partial_{\phi}\Upsilon^{1}(\phi)\ ,\quad \eta^{2}(u,\phi)=S(\phi)+u\partial_{\phi}\epsilon^{2}(\phi)\ .
\end{align}
Finally, the gauge variation equations also provide us with the variation of the 
independent functions $\mathcal{M}, \mathcal{L}, \mathcal{U}$ and $\mathcal{Q}$ appearing in the asymptotic form of the metric in terms of the independent gauge parameters $\Upsilon^1(\phi), \epsilon^2(\phi), T(\phi)$ and $S(\phi)$. These are given by
\begin{eqnarray} \label{fv}
   \delta \mathcal{M} &=& -2\partial_{\phi}^{3} \Upsilon^{1} +  \partial_{\phi} \mathcal{M}.\Upsilon^{1}+2\mathcal{M}.\partial_{\phi} \Upsilon^{1} - 32\sigma \partial_{\phi} \mathcal{Q}.\epsilon^{2}-48\sigma \mathcal{Q} \partial_{\phi} \epsilon^{2}\ ,\\
   \delta \mathcal{L} &=& -2\partial_{\phi}^{3} T + \partial_{\phi} \mathcal{M} .T+2\mathcal{M}.\partial_{\phi} T +\partial_{\phi} \mathcal{L}. \Upsilon^{1}+2\mathcal{L}.\partial_{\phi} \Upsilon^{1} \nonumber \\
   &&\hspace*{2.5cm} - 32\sigma \partial_{\phi} \mathcal{Q}. S -48\sigma \mathcal{Q} .\partial_{\phi} S- 32\sigma \partial_{\phi} \mathcal{U}.\epsilon^{2}-48\sigma \mathcal{U}. \partial_{\phi} \epsilon^{2}\\
   \delta \mathcal{Q} &=& \frac{1}{24} \partial_{\phi}^{5} \epsilon^{2} -\frac{1}{24}\partial_{\phi}^{3} \mathcal{M}. \epsilon^{2}-\frac{3}{16}\partial_{\phi}^{2}\mathcal{M}.\partial_{\phi} \epsilon^{2}-\frac{5}{16}\partial_{\phi}\mathcal{M}.\partial_{\phi}^{2} \epsilon^{2}-\frac{5}{24}\mathcal{M}.\partial_{\phi}^{3} \epsilon^{2}\nonumber\\
    &&\hspace*{2.5cm}+\frac{1}{6}\mathcal{M}.\partial_{\phi}\mathcal{M}.\epsilon^{2}+\frac{1}{6}\mathcal{M}^{2}.\partial_{\phi}\epsilon^{2}+\partial_{\phi}\mathcal{Q}.\Upsilon^{1}+3\mathcal{Q}.\partial_{\phi}\Upsilon^{1} \\
    \delta \mathcal{U}&=& \frac{1}{24} \partial_{\phi}^{5}S -\frac{1}{24}\partial_{\phi}^{3} \mathcal{M}.S-\frac{3}{16}\partial_{\phi}^{2}\mathcal{M}.\partial_{\phi} S-\frac{5}{16}\partial_{\phi}\mathcal{M}.\partial_{\phi}^{2} S-\frac{5}{24}\mathcal{M}.\partial_{\phi}^{3} S\nonumber\\
&&+\frac{1}{6}\mathcal{M}.\partial_{\phi}\mathcal{M}.S+\frac{1}{6}\mathcal{M}^{2}.\partial_{\phi}S-\frac{1}{24}\partial_{\phi}^{3} \mathcal{L}. \epsilon^{2}-\frac{3}{16}\partial_{\phi}^{2}\mathcal{L}.\partial_{\phi} \epsilon^{2}-\frac{5}{16}\partial_{\phi}\mathcal{L}.\partial_{\phi}^{2} \epsilon^{2}\nonumber\\
&&-\frac{5}{24}\mathcal{L}.\partial_{\phi}^{3} \epsilon^{2}+\frac{1}{6}\partial_{\phi}\mathcal{M}.\mathcal{L}.\epsilon^{2}+\frac{1}{6}\mathcal{M}.\partial_{\phi}\mathcal{L}.\epsilon^{2}+\frac{1}{3}\mathcal{M}.\mathcal{L}.\partial
_{\phi}\epsilon^{2}+\partial_{\phi}\mathcal{Q}T\nonumber\\
&&\hspace*{4cm}+3\mathcal{Q}\partial_{\phi}T+\partial_{\phi}\mathcal{U}\Upsilon^{1}+3\mathcal{U}\partial_{\phi}\Upsilon^{1}\ .
\end{eqnarray}

For a generic Chern-Simons theory, using the canonical approach, the variation of the canonical generators that corresponds to the asymptotic symmetries of this theory can be found as,
\begin{align}
    \delta Q = -\frac{k}{2\pi} \int \langle \Lambda, \delta A_{\phi} \rangle d\phi
\end{align}

In our present case, using appropriate supertrace elements, the above expression of variation of the charge reduces to,
\begin{align}
    \delta Q=-\frac{k}{2\pi} \int \left[\frac{1}{8} \Upsilon^{1} \delta\mathcal{N}+\frac{1}{8} \xi^{1} \delta\mathcal{M}+2\sigma \epsilon^{2} \delta\mathcal{O}+2
    \sigma \eta^{2} \delta\mathcal{Q}\right]~d\phi
\end{align}

We need to integrate this relation to find the asymptotic charge. For this purpose, the expressions must be expressed in terms of independent gauge variation parameters, whose variations are set to zero values at the boundary. Writing the variation of the charge in terms 
of independent fields $\mathcal{L}, \mathcal{M}, \mathcal{U}, \mathcal{Q}$  and gauge parameters $\Upsilon^{1}, T, \epsilon^{2}, S$. The variation  of the charge is given as,
\begin{align}
    \delta Q=-\frac{k}{2\pi} \int \left[\frac{1}{8} \Upsilon^{1} \delta\mathcal{L}+\frac{1}{8} T \delta\mathcal{M}+2\sigma \epsilon^{2} \delta\mathcal{U}+2
    \sigma S \delta\mathcal{Q}\right]~d\phi
\end{align}

We now easily find the expression of the charge under some mild regularity assumptions of variations and it is given as,
\begin{align}
     Q=-\frac{k}{2\pi} \int \left[\frac{1}{8} \Upsilon^{1} \mathcal{L}+\frac{1}{8} T \mathcal{M}+2\sigma \epsilon^{2} \mathcal{U}+2
    \sigma S \mathcal{Q}\right]~d\phi .
\end{align}

A further re-scaling of fields, as suggested in \cite{Afshar:2013vka} is useful in bringing the above charge into a more suitable form. It is given as,
\begin{align}
    \tilde{\mathcal{L}}=\frac{k}{4\pi}\mathcal{L};~\tilde{\mathcal{M}}=\frac{k}{4\pi}\mathcal{M};~\tilde{\mathcal{U}}=\frac{k}{\pi}\mathcal{U};~\tilde{\mathcal{Q}}=\frac{k}{\pi}\mathcal{Q};~\tilde{\epsilon^{2}}=576 \epsilon^{2};~\tilde{S}=576 S
\end{align}
Using the above scaling, the final expression of the asymptotic charge can be achieved and is given as,
\begin{align}
    Q=-\int \left[\Upsilon^{1} \tilde{\mathcal{L}}+ T \tilde{\mathcal{M}}+\frac{\sigma}{144} \tilde{\epsilon}^{2} \tilde{\mathcal{U}}+\frac{\sigma}{144}\tilde{S} \tilde{\mathcal{Q}}\right]~d\phi
\end{align}

The asymptotic symmetry algebra can be found using asymptotic charge and its variation. Precisely, the Poisson brackets among various modes of the fields can be found using the formula,
\begin{align}\label{caf}
    \{Q[\lambda_{1}],Q[\lambda_{2}]\}_{P.B.}=\delta_{\lambda_{1}} Q[\lambda_{2}]
\end{align}

   We have the expression of the charge in terms of independent fields and parameters and using the above formula \eqref{caf} and the variation of different fields \eqref{fv}, one can now write the full asymptotic algebra. It is important to note that, while using the field variations, we have split the quadratic terms involving different fields in explicit symmetric combinations. The reason for the same will be clarified in the next subsection. The complete asymptotic algebra for a gravity theory connected to spin-3 is given as,
    \begin{align}\label{chargealgebra}
        i\{\mathcal{L}_{m},\mathcal{L}_{n}\}=&(m-n)\mathcal{L}_{m+n} \nonumber\\
     i\{\mathcal{L}_{m},\mathcal{M}_{n}\}=&(m-n)\mathcal{M}_{m+n}+\frac{c_{M}}{12} m(m^{2}-1) \delta_{m+n,0} \nonumber\\
        i\{\mathcal{L}_{m},\mathcal{U}_{n}\}=&(2m-n)\mathcal{U}_{m+n} \nonumber\\
       i\{\mathcal{L}_{m},\mathcal{V}_{n}\}=&(2m-n)\mathcal{V}_{m+n} \nonumber\\
       i\{\mathcal{M}_{m},\mathcal{U}_{n}\}=&(2m-n)\mathcal{V}_{m+n} \nonumber\\
     i\{\mathcal{U}_{m},\mathcal{U}_{n}\}=&\frac{3}{\sigma} \bigg[(m-n)(2m^{2}+2n^{2}-mn-8)\mathcal{L}_{m+n}\bigg. \nonumber\\
     &\hspace*{3cm}\bigg. +\frac{96}{c_{M}}(m-n)\sum_{p}(\mathcal{M}_{m+n-p} \mathcal{L}_{p}+\mathcal{L}_{m+n-p} \mathcal{M}_{p})\bigg]\nonumber\\
     i\{\mathcal{U}_{m},\mathcal{V}_{n}\}=&\frac{3}{\sigma} \bigg[(m-n)(2m^{2}+2n^{2}-mn-8)\mathcal{M}_{m+n}+\frac{96}{c_{M}}(m-n)\sum_{p}\mathcal{M}_{m+n-p}\mathcal{M}_{p}\nonumber\\
     &\hspace*{3cm}+\frac{c_{M}}{12} m (m^{2}-1)(m^{2}-4)  \delta_{m+n,0} \bigg]\ ,
    \end{align}
 where $c_{M}=12k$ and we have defined the Fourier modes as,
\begin{align}
\tilde{\mathcal{L}}(\phi)=&\frac{1}{2\pi}\sum_{m} \mathcal{L}_{m} e^{-im\phi},~\tilde{\mathcal{M}}(\phi)=\frac{1}{2\pi}\sum_{m} \mathcal{M}_{m} e^{-im\phi},\nonumber\\
\tilde{\mathcal{U}}(\phi)=&\frac{1}{2\pi}\sum_{m} \mathcal{U}_{m} e^{-im\phi},~\tilde{\mathcal{Q}}(\phi)=\frac{1}{2\pi}\sum_{m} \mathcal{V}_{m} e^{-im\phi}.
\end{align}

In the above, 
we have made a shift in the zero mode $\mathcal{M}_{0} \rightarrow \mathcal{M}_{0}+\frac{c_{M}}{24}$ to make sure to get the centerless global algebra as sub algebra of asymptotic algebra for $\{\mathcal{L}_{m},\mathcal{M}_{m}\}$ where $m=0,\pm 1$ and $\{\mathcal{U}_{m},\mathcal{V}_{m}\}$ where $m=0,\pm 1,\pm 2$. Thus we have identified the asymptotic symmetry algebra of spin 3 extended gravity theory, for an arbitrary value of the spin 3 coupling parameter $\sigma$. This is one of the results of this paper. In the next section, we derive the same algebra by using Inn\"on\"u-Wigner Contraction of two copies of spin 3 extended asymptotic AdS algebra. 

\subsection{Alternate way of deriving the algebra : Inn\"on\"u-Wigner Contraction}
The Inn\"on\"u-Wigner contraction is a group contraction act on a Lie algebra to get a different Lie algebra which is, in general, non-isomorphic to the previous Lie algebra with respect to a continuous subgroup of it. The contraction (limiting) operation on a parameter of the Lie algebra under suitable conditions alters the structure constants of the corresponding Lie algebra in a non-trivial singular way. The same idea of Inn\"on\"u-Wigner contraction has been used to get the desired asymptotic algebra for the flat (super)gravity theories from the asymptotic algebra of the corresponding  AdS case by taking a suitable limit. In the present case, one can also derive the charge algebra \eqref{chargealgebra} from a suitable combination of two copies of asymptotic higher spin charge algebra of AdS spacetime by considering the AdS radius $l\rightarrow \infty$ limit. One can start with two copies of $W_{3}$ algebra \cite{Campoleoni:2010zq}
\begin{align}
    &i\{\mathcal{J}_{p},\mathcal{J}_{q}\}= (p-q)\mathcal{J}_{p+q}+\frac{c}{12}p(p^{2}-1)\delta_{p+q,0} \nonumber\\
    &i\{\mathcal{J}_{p},\mathcal{W}_{q}\}= (2p-q)\mathcal{W}_{p+q} \nonumber\\
    &i\{\mathcal{W}_{p},\mathcal{W}_{q}\}= -\frac{\sigma}{3}\left[(p-q)(2p^{2}+2q^{2}-pq-8)\mathcal{J}_{p+q}+\frac{96}{c}(p-q)\Lambda_{p+q}\right. \nonumber\\
    &\hspace*{8.5cm}\left.+\frac{c}{12}p(p^{2}-1)(p^{2}-4)\delta_{p+q,0}\right] \nonumber\\
    &i\{\bar{\mathcal{J}}_{p},\bar{\mathcal{J}}_{q}\}= (p-q)\bar{\mathcal{J}}_{p+q} +\frac{\bar{c}}{12}p(p^{2}-1)\delta_{p+q,0}\nonumber\\
    &i\{\bar{\mathcal{J}}_{p},\bar{\mathcal{W}}_{q}\}= (2p-q)\bar{\mathcal{W}}_{p+q} \nonumber\\
    &i\{\bar{\mathcal{W}}_{p},\bar{\mathcal{W}}_{q}\}= -\frac{\sigma}{3}\left[ (p-q)(2p^{2}+2q^{2}-pq-8)\bar{\mathcal{J}}_{p+q}+\frac{96}{\bar{c}}(p-q)\bar{\Lambda}_{p+q}\right.\nonumber\\
    &\hspace*{8.5cm}\left.+\frac{\bar{c}}{12}p(p^{2}-1)(p^{2}-4)\delta_{p+q,0} \right] \nonumber
\end{align}
where, $\Lambda_{p}=\sum_{q \in \mathbb{Z}} \mathcal{J}_{p+q} \mathcal{J}_{-q} $, $\bar{\Lambda}_{p}=\sum_{q \in \mathbb{Z}} \bar{\mathcal{J}}_{p+q} \bar{\mathcal{J}}_{-q}$ and  $c=\bar{c}=\frac{3l}{2G}$.

The algebra for the flat case can be obtained by introducing the singular map between the generators of the two copies of AdS algebras and the generators of the flat algebra. The relationship between generators of AdS algebra $\{\mathcal{J}_{n},\bar{\mathcal{J}}_{n},\mathcal{W}_{n},\bar{\mathcal{W}}_{n}\}$  and the generators of flat algebra $\{\mathcal{L}_{n},\mathcal{M}_{n},\mathcal{U}_{n},\mathcal{V}_{n}\}$ is given as,
\begin{align}
    \mathcal{L}_{n}=\mathcal{J}_{n}-\bar{\mathcal{J}}_{-n},~\mathcal{M}_{n}=\frac{\mathcal{J}_{n}+\bar{\mathcal{J}}_{-n}}{l},~\mathcal{U}_{n}=\mathcal{W}_{n}-\bar{\mathcal{W}}_{-n},~\mathcal{V}_{n}=\frac{\mathcal{W}_{n}+\bar{\mathcal{W}}_{-n}}{l}
\end{align}
With the above mapping among generators, we can write the flat algebra in terms of generators $\{\mathcal{L}_{n},\mathcal{M}_{n},\mathcal{U}_{n},\mathcal{V}_{n}\}$. The non-zero commutators of the algebra in the form of  Poisson bracket can be written as,

 \begin{align}\label{chargealgebra1}
        i\{\mathcal{L}_{m},\mathcal{L}_{n}\}=&(m-n)\mathcal{L}_{m+n}\nonumber\\
     i\{\mathcal{L}_{m},\mathcal{M}_{n}\}=&(m-n)\mathcal{M}_{m+n}+\frac{c_{M}}{12} m(m^{2}-1) \delta_{m+n,0} \nonumber\\
        i\{\mathcal{L}_{m},\mathcal{U}_{n}\}=&(2m-n)\mathcal{U}_{m+n} \nonumber\\
       i\{\mathcal{L}_{m},\mathcal{V}_{n}\}=&(2m-n)\mathcal{V}_{m+n} \nonumber\\
       i\{\mathcal{M}_{m},\mathcal{U}_{n}\}=&(2m-n)\mathcal{V}_{m+n} \nonumber\\
     i\{\mathcal{U}_{m},\mathcal{U}_{n}\}=&-\frac{\sigma}{3}\bigg[(m-n)(2m^{2}+2n^{2}-mn-8)\mathcal{L}_{m+n} \bigg. \nonumber\\
     &\hspace*{3cm}\bigg.+\frac{96}{c_{M}}(m-n)\sum_{k}(\mathcal{M}_{m+n-p}\mathcal{L}_{p}+\mathcal{L}_{m+n-p}\mathcal{M}_{p}
     )\bigg]\nonumber\\
     i\{\mathcal{U}_{m},\mathcal{V}_{n}\}=&-\frac{\sigma}{3}\bigg[(m-n)(2m^{2}+2n^{2}-mn-8)\mathcal{M}_{m+n}+\frac{96}{c_{M}}(m-n)\sum_{k}\mathcal{M}_{m+n-p}\mathcal{M}_{p}\nonumber\\
     &\hspace*{3cm}+\frac{c_{M}}{12} m (m^{2}-1)(m^{2}-4)  \delta_{m+n,0}\bigg]
    \end{align}
    where finally we have taken flat limit $l \rightarrow \infty$. We have further identified $c_{M}=\frac{c+\bar{c}}{l}=\frac{2c}{l}=\frac{3}{G}$. The Poisson Brackets exactly match with the ones obtained in \eqref{chargealgebra}. The point to note is that the Inn\"on\"u-Wigner contraction of the two copies of the spin-3 AdS algebras clearly justifies the symmetrization of non-identical fields in the gauge variation that we had performed in the last subsection. In other words, while reviewing the asymptotic symmetry algebra of higher spin 3 fields coupled to gravity in terms of that of the asymptotically AdS theory, in the large $l$ limit the fields appear in a particular symmetric combination. Further one can obtain the most generic possible quantum extension of the above Poisson Bracket relations by allowing the central term in $[\mathcal{L}_{m},\mathcal{L}_{n}]$ and $[\mathcal{U}_{m},\mathcal{U}_{n}]$ commutator as presented in \cite{Afshar:2013vka}. Note that the level of the corresponding Chern-Simon actions is related as $k_{l}=k.l$.


\section{The Dual Theory }\label{sec5}
In the last section, we obtained the asymptotic symmetry algebra for three dimensional asymptotically flat gravity with a spin 3 generalization. Now our main objective is to construct a two dimensional theory at the boundary which is dual to our higher spin gravity theory in the bulk. This is where the Chern-Simons formulation of (2+1) dimensional gravity plays a crucial role. It is well known that Chern-Simons theory defined on a manifold with boundaries is equivalent to a WZW theory with appropriate boundary conditions. Although WZW theory has a three dimensional piece, its dynamics are effectively two dimensional and are restricted to the boundary of the original CS theory. This can be better understood by further reducing the WZW theory to a Liouville type theory defined exclusively at the 2D boundary.

Since the effect of boundary is crucial in this correspondence between different theories, let us discuss it in a bit more detail. For this, we will split the CS Gauge field into two parts $A = du\, A_u + \tilde{A}$ where $A_u$ is the timelike component and $\Tilde{A}$ is the spatial part of the field. Then we can rewrite the CS action in Hamiltonian form (See Appendix B of \cite{Banerjee:2019lrv} for details of the derivation) as,
\begin{align}
    I_H [A] = k \int \langle \Tilde{A},~ du \dot{\Tilde{A}} \rangle + 2 \langle du A_u, ~\Tilde{d}\Tilde{A}+ \Tilde{A}^2 \rangle
\end{align}
where the derivatives are also split into the temporal and the spatial parts. Now from this form of the action, it becomes clear that the variation of the action is not well defined for a generic boundary condition. In fact, the variation of the action yields an extra boundary term of the form $-2k \int du\,\tilde{d} \langle A_u, \delta \tilde{A} \rangle $. This can be further simplified in our case since the spatial directions are $\{r,\phi\}$ and $\phi$ is a compact direction. Hence, the only surviving boundary term is in $(u,\phi)$ surface.

Since CS is a topological field theory, it is locally trivial. This is also clear from the fact that the equation of motion says that the field strength vanishes at every point. Hence, locally, the solution for the gauge connection can be written as 
\begin{equation}
    A_i=G^{-1}\partial_i G
\end{equation}
where G is an element of the gauge group. We further impose gauge choice to write $$G (u,r,\phi) = g(u,\phi) h(u,r).$$ Then following our argument an improved CS action along with the proper boundary term takes the form: 
\begin{equation}\label{wzwg}
     I_{\text{CS}}[g]=-k\int_{\partial \mathcal{M}} d^{2}x \epsilon^{ij}~ Tr( g^{-1}\partial_{i}g g^{-1}\partial_{j}g ) +\frac{k}{3}\int_{\mathcal{M}} d^3x \epsilon^{\mu\nu\rho}~\text{Tr}(G^{-1}\partial_{\mu}G G^{-1}\partial_{\nu}GG^{-1}\partial_{\rho}G) + I_{\text{bdy}}
\end{equation}\\
which is the chiral Wess-Zumino-Witten action. The variation of this action is two dimensional which already manifests the 2D nature of our dual holographic theory. We can further reduce this chiral WZW action to a Liouville type theory. As we will show, doing so requires us to use the full set of boundary conditions on $A$ that we obtained from the gravitational context. In the following, our main objective would be to identify such a theory for the higher spin 3 gravity case.

\subsection{Action for a three dimensional flat gravity theory connected to spin-3 field}

Starting from \eqref{wzwg}, the Wess-Zumino-Witten action with the coordinate choice of $(u,\phi,r)$ for the bulk space time can be expressed as,
\begin{align}\label{wzw}
    I_{CS}[g]=k\int_{\partial \mathcal{M}} du d\phi \langle \partial_{\phi} g g^{-1} \partial_{u}g g^{-1}\rangle + \frac{k}{3}\int_{\mathcal{M}} \langle G^{-1}dG \wedge G^{-1}dG \wedge G^{-1}dG \rangle + I_{bdy}.
\end{align}
 In the above action, $g$ is the dynamical field living on the boundary spanned by $(u,\phi)$ and $G$ is its pullback to the bulk manifold. To evaluate this action we use Gauss decomposition for the field $g$. Further, we write the above action \eqref{wzw} in a more suitable form with the introduction of Polyakov-Wiegmann identity \cite{Polyakov:1984et} \footnote{
  look at the appendix A for relevant details}. In the present case, splitting $g=g_1 g_2 g_3$, we express the action \eqref{wzw} as,
\begin{align}  \label{pwia}
I_{CS}(g)=&I(g_{1})+I(g_{2})+I(g_{3})+2k\int_{\partial\mathcal{M}} du d\phi \text{Tr}(g_{1}^{-1}\partial_{\phi} g_{1}\ \partial_{u} g_{2} g_{2}^{-1}) \nonumber\\
    &+2k\int_{\partial\mathcal{M}} du d\phi \text{Tr}(g_{2}^{-1}\partial_{\phi} g_{2}\ \partial_{u} g_{3} g_{3}^{-1})+2k\int_{\partial\mathcal{M}} du d\phi \text{Tr}(g_{1}^{-1}\partial_{\phi} g_{1}\ g_{2}\partial_{u} g_{3} g_{3}^{-1} g_{2}^{-1})\nonumber\\
    &\hspace*{11.5cm}+I_{bdy}
\end{align}
 For performing the Gauss decomposition\footnote{Important aspects of Gauss decomposition are discussed in \cite{Tsutsui:1994pp}} of fields, we need to expand the fields into the Chevalley-Serre basis of the corresponding gauge group. In the present case, the gauge group is $\mathfrak{fhs}(3)$ as given in \eqref{HSBA}. The decomposition is given as,
 \begin{align}\label{gd}
    g_{1}&=e^{X L_{1}+W M_{1}+P U_{1}+Q U_{2}+R V_{1}+S V_{2}}\nonumber\\
     g_{2}&=e^{\Phi L_{0}+\zeta M_{0}+\eta U_{0}+\xi V_{0}}\nonumber\\
     g_{3}&=e^{Y L_{-1}+V M_{-1}+E U_{-1}+F U_{-2}+C V_{-1}+D V_{-2}}
\end{align}

where $\{L_{i},M_{i}\}(i=0,\pm 1)$ and $\{U_{i},V_{i}\}(i=0,\pm 1,\pm 2)$ are the generators of $\mathfrak{fhs}(3)$ \eqref{HSBA}. Note that $X, W, P, Q, R, S, \Phi, \zeta, \eta, \xi, Y, V, E, F, C, D$ are the boundary fields and are functions of $u, \phi$ coordinate only. Using appropriate matrix representation\footnote{look at appendix C} of the generators where the matrices are upper triangular, diagonal and lower triangular matrices respectively, from \eqref{gd} we find different terms of the action
\eqref{pwia}. $I(g_{1}),  I(g_{2}),  I(g_{3})$ are given as,
 \begin{align}
     I(g_{1})&=0 \\
     I(g_{2})&=k \int_{\partial\mathcal{M}} du d\phi~\Big(\frac{1}{4}\Phi^{\prime}\dot{\zeta}+\frac{1}{4}\zeta^{\prime}\dot{\Phi}+\frac{\sigma}{3}\xi^{\prime}\dot{\eta}+\frac{\sigma}{3}\eta^{\prime}\dot{\xi} \Big)\\
     I(g_{3})&=0
 \end{align}
 and the rest of the three pieces can be found in the same way ({see appendix for detailed expressions}).
 Putting together all the pieces, the full action \eqref{pwia} can be expressed as,
 \begin{align}\label{Bdyaction}
 &I_{CS}= k \int_{\partial\mathcal{M}} du d\phi \Big[ \frac{1}{4}\Phi^{\prime}\dot{\zeta}+\frac{1}{4}\zeta^{\prime}\dot{\Phi}+\frac{\sigma}{3}\xi^{\prime}\dot{\eta}+\frac{\sigma}{3}\eta^{\prime}\dot{\xi} \nonumber\\
 &+ 2\Big( \frac{1}{2} \sigma e^{2\Phi} (V X+ WY) P^{\prime} \dot{E} -\frac{1}{2}\sigma \cosh({2\sqrt{\sigma}\eta}) e^{\Phi} \zeta P^{\prime} \dot{E}+\sigma e^{2\Phi} X Y \zeta  P^{\prime} \dot{E} \nonumber\\
      &-\sigma^{\frac{3}{2}} \cosh({2\sqrt{\sigma} \eta}e^{\Phi}) \xi P^{\prime} \dot{E}+\sigma e^{2\Phi} (V -2Y \zeta) Q^{\prime}\dot{E}-\frac{1}{2} \sigma \cosh({2\sqrt{\sigma} \eta})e^{\Phi}R^{\prime} \dot{E}+\frac{1}{2}\sigma e^{2\Phi} X Y R^{\prime} \dot{E}\nonumber\\
      &-\sigma e^{2\Phi} Y S^{\prime} \dot{E}+\frac{1}{2} \sqrt{\sigma} \cosh({2\sqrt{\sigma} \eta}) e^{\Phi}W^{\prime} \dot{E}-\frac{1}{2} \sigma e^{2\Phi} P Y W^{\prime} \dot{E}-\frac{1}{2} \sigma e^{2\Phi} (P V+R Y) X^{\prime} \dot{E}\nonumber\\
      &+\frac{1}{2} \sqrt{\sigma} \cosh({2\sqrt{\sigma} \eta})e^{\Phi} \zeta X^{\prime} \dot{E} -\sigma e^{2\Phi} P Y \zeta X^{\prime} \dot{E}- \sigma \cosh({2\sqrt{\sigma}} \eta)e^{\Phi} \xi X^{\prime} \dot{E}-\sigma e^{2\Phi} W P^{\prime} \dot{F}\nonumber\\
      &-2\sigma e^{2\Phi} X \zeta P^{\prime} \dot{F}+4\sigma e^{2\Phi} \zeta Q^{\prime} \dot{F}-\sigma e^{2\Phi} X R^{\prime} \dot{F}+2\sigma e^{2\Phi} S^{\prime} \dot{F}+\sigma e^{2\Phi} P W^{\prime} \dot{F}+\sigma e^{2\Phi} R X^{\prime} \dot{F}\nonumber\\
      &+2\sigma e^{2\Phi} P \zeta X^{\prime} \dot{F}-\frac{1}{2} \sigma \cosh({2\sqrt{\sigma} \eta})e^{\Phi}P^{\prime} \dot{C}+\frac{1}{2} \sigma e^{2\Phi} X Y P^{\prime} \dot{C}-\sigma e^{2\Phi} Y Q^{\prime} \dot{E}\nonumber\\
      &+\frac{1}{2} \sqrt{\sigma} \cosh({2\sqrt{\sigma} \eta})e^{\Phi}X^{\prime} \dot{C}-\frac{1}{2} \sigma e^{2\Phi} P Y X^{\prime} \dot{C}-\sigma e^{2\Phi} X P^{\prime} \dot{D}-2\sigma e^{2\Phi} Q^{\prime} \dot{D}\nonumber\\
      &+\sigma e^{2\Phi} P X^{\prime} \dot{D}+\frac{1}{2} \sqrt{\sigma} \sinh({2\sqrt{\sigma} \eta})e^{\Phi}P^{\prime} \dot{V}-\frac{1}{2} \sigma e^{2\Phi} E X P^{\prime} \dot{V}+\sigma e^{2\Phi} E Q^{\prime} \dot{V}\nonumber\\
      &-\frac{1}{2} \sinh({2\sqrt{\sigma} \eta})e^{\Phi}X^{\prime} \dot{V}+\frac{1}{2} \sigma e^{2\Phi} E P X^{\prime} \dot{V}-\frac{1}{2} \sigma e^{2\Phi} E W P^{\prime} \dot{Y}-\frac{1}{2} \sigma e^{2\Phi} C X P^{\prime} \dot{Y}\nonumber\\
      &+\frac{1}{2} \sqrt{\sigma} \sinh({2\sqrt{\sigma} \eta})e^{\Phi} \zeta P^{\prime} \dot{Y}-\sigma e^{2\Phi} E X \zeta P^{\prime} \dot{Y}\- \sigma \cosh({2\sqrt{\sigma} \eta})e^{\Phi} \xi P^{\prime} \dot{Y}+\frac{1}{2}\sigma e^{2\Phi} E P  W^{\prime} \dot{Y}\nonumber\\
      &+ \sigma e^{2\Phi} C  Q^{\prime} \dot{Y}+ 2\sigma e^{2\Phi} E \zeta  Q^{\prime} \dot{Y}+\frac{1}{2} \sqrt{\sigma} \sinh({2\sqrt{\sigma} \eta})e^{\Phi}  R^{\prime} \dot{Y}-\frac{1}{2} \cosh({2\sqrt{\sigma} \eta})e^{\Phi}W^{\prime} \dot{Y} \nonumber\\
      &+\frac{1}{2}\sigma e^{2\Phi} C P  X^{\prime} \dot{Y}+\frac{1}{2}\sigma e^{2\Phi} E R  X^{\prime} \dot{Y}-\frac{1}{2} \cosh({2\sqrt{\sigma} \eta})e^{\Phi} \zeta X^{\prime} \dot{Y}+\sigma e^{2\Phi} E P \zeta  X^{\prime} \dot{Y}\nonumber\\
      &+\sqrt{\sigma} \sinh({2\sqrt{\sigma} \eta})e^{\Phi} \xi X^{\prime} \dot{Y}\Big)\Big]+ I_{bdy}
 \end{align}

The above action contains all the fields that appeared in the Gauss Decomposition of the gauge field. This is certainly an overestimation, as to get the boundary theory that sees \eqref{chargealgebra} as its symmetry, we further need to use the constraints on various fields coming from the appropriate boundary condition \eqref{BC}.  The boundary conditions \eqref{BC} actually tells us that the fields introduced above are not all independent, rather they depend on each other at the boundary. The boundary condition imposes the following constraints on the fields,
\begin{align}\label{CR}
   &Q^{\prime}=\frac{1}{2}(XP^{\prime}-PX^{\prime}),\nonumber\\
   &~\sqrt{\sigma} e^{\Phi}P^{\prime}=-\sinh({2\sqrt{\sigma} \eta}),~~e^{\Phi}X^{\prime} = \cosh({2\sqrt{\sigma} \eta})\nonumber\\
   &E=-\frac{\eta^{\prime}}{3},~Y=-\frac{\Phi^{\prime}}{2},~C=-\frac{\xi^{\prime}}{3},~Y=-\frac{\zeta^{\prime}}{2},~F=\frac{1}{6}\eta^{\prime}\Phi^{\prime}+\frac{1}{12}\eta^{\prime\prime}\nonumber\\
   &\sqrt{\sigma}e^{\Phi}R^{\prime}=\sinh({2\sqrt{\sigma} \eta})\zeta-2\sqrt{\sigma}\cosh({2\sqrt{\sigma} \eta})\xi,\nonumber\\
   &~e^{\Phi}W^{\prime} = -\cosh({2\sqrt{\sigma} \eta})\zeta+2\sqrt{\sigma} \sinh({2\sqrt{\sigma} \eta})\xi \nonumber\\
   &S^{\prime}=\frac{1}{2}(WP^{\prime}+XR^{\prime}-PW^{\prime}-RX^{\prime}),\nonumber\\
   &~D=\frac{1}{6}\zeta^{\prime}\eta^{\prime}+\frac{1}{6}\xi^{\prime}\Phi^{\prime}+\frac{1}{12}\xi^{\prime\prime}
\end{align}
Thus we see that we have 12 constraint relations among 16 fields and hence there will be only 4 independent degrees of freedom $\Phi,\zeta,\eta,\xi$ for the boundary field theory. These independent fields are related to the four independent components $\mathcal{M},\mathcal{N},\mathcal{Q},\mathcal{O}$ of the asymptotic Chern-Simons field which can be realized from the boundary conditions as follows,
\begin{equation}
 \begin{aligned}\label{FR}
     \mathcal{M}&=\frac{4\sigma}{3}(\eta^{\prime})^{2}+(\Phi^{\prime})^{2}+2\Phi^{\prime\prime}\nonumber\\
     \mathcal{N}&=\frac{8\sigma}{3}\eta^{\prime}\xi^{\prime}+2\zeta^{\prime}\Phi^{\prime}+2\zeta^{\prime\prime}\nonumber\\
     \mathcal{Q}&=-\frac{2\sigma}{27} (\eta^{\prime})^{3}+\frac{1}{6}\eta^{\prime}(\Phi^{\prime})^{2}+\frac{1}{4}\Phi^{\prime}\eta^{\prime\prime}+\frac{1}{12}\eta^{\prime}\Phi^{\prime\prime}+\frac{1}{12}\eta^{\prime\prime\prime}\nonumber\\
     \mathcal{O}&=-\frac{5\sigma}{18}(\eta^{\prime})^{2}\xi^{\prime}+\frac{1}{3}\zeta^{\prime}\eta^{\prime}\Phi^{\prime}+\frac{1}{8}\xi^{\prime}(\Phi^{\prime})^{2}+\frac{1}{12}\eta^{\prime}\zeta^{\prime\prime}+\frac{1}{4}\zeta^{\prime}\eta^{\prime\prime}+\frac{7}{24}\Phi^{\prime}\xi^{\prime\prime}+\frac{1}{24}\xi^{\prime}\Phi^{\prime\prime}+\frac{1}{12}\xi^{\prime\prime\prime}
 \end{aligned}
\end{equation}
Plugging all the expression from \eqref{CR} and \eqref{FR}, The dual action in  \eqref{Bdyaction} can finally be written in terms of the independent fields $\Phi, \zeta, \eta, \xi$ as,
\begin{align}\label{pra}
    I_{CS}=k \int_{\partial\mathcal{M}} du d\phi \Big[\frac{1}{4}\Phi^{\prime}\dot{\zeta}+\frac{1}{4}\zeta^{\prime}\dot{\Phi}+\frac{\sigma}{3}\xi^{\prime}\dot{\eta}+\frac{\sigma}{3}\eta^{\prime}\dot{\xi}+\frac{1}{2}\dot{\zeta}^{\prime}\Big] + I_{bdy}
\end{align}

Here we have expressed the action in terms of the independent fields up to the boundary term. 
The boundary term contributes as the components of the gauge field satisfy a particular boundary condition \eqref{BC}. The generic form of the boundary term, in terms of the gauge field components, can be found by demanding a well defined variation of the action and is given as, 
\begin{align}\label{varbdy}
    \delta I_{bdy}=-4k\int_{\mathcal{M}} du d\phi dr~\langle \partial_{r}(A_{u}\delta A_{\phi})-\partial_{\phi}(A_{u}\delta A_{r}) \rangle
\end{align}

Boundary coordinate $\phi$ is cyclic in nature, and we will get no contribution from the term involving a total derivative with respect to $\phi$.  Thus \eqref{varbdy} reduces to,
\begin{align}
   \delta I_{bdy}=-4k\int_{\mathcal{M}} du d\phi dr~\langle \partial_{r}(A_{u}\delta A_{\phi}) \rangle
\end{align}

After radial integration, we can take out the variation and can express the boundary term as follows,
\begin{align}
   I_{bdy}=-2k\int_{\partial \mathcal{M}} du d\phi ~\langle A_{u} A_{\phi} \rangle
\end{align}
To evaluate this boundary term we have to use \eqref{BC} and the information of the supertrace elements. Using these, the boundary term takes the form given as,
\begin{align}
   I_{bdy}=-2k\int_{\partial \mathcal{M}} du d\phi ~\Big(\frac{\mathcal{M}}{4} \Big)
\end{align}

We can further rewrite the boundary action in terms of the independent fields of the 2D dual field theory using \eqref{FR} as,
\begin{align}\label{bt}
    I_{bdy}=-2k\int_{\partial\mathcal{M}} du d\phi \Big[\frac{\sigma}{3}(\eta^{\prime})^{2}+\frac{1}{4}(\Phi^{\prime})^{2}+\frac{1}{2}\Phi^{\prime\prime} \Big] .
\end{align}

 Incorporating the above boundary action in \eqref{pra}, we get the following reduced dual action,
\begin{align}
     I_{\text{CS}}=k \int_{\partial\mathcal{M}} du d\phi \Big[\frac{1}{4}\Phi^{\prime}\dot{\zeta}+\frac{1}{4}\zeta^{\prime}\dot{\Phi}+\frac{\sigma}{3}\xi^{\prime}\dot{\eta}+\frac{\sigma}{3}\eta^{\prime}\dot{\xi}+\frac{1}{2}\dot{\zeta}^{\prime}-\frac{2\sigma}{3}(\eta^{\prime})^{2}-\frac{1}{2}(\Phi^{\prime})^{2}-\Phi^{\prime\prime}\Big]
\end{align}
The compact nature of angular direction $\phi$ will help us to manipulate some terms. After simplification,  we can write the final form of the reduced action as,
\begin{align}\label{eq:final-2Dtheory}
     I_{\text{CS}}=k \int_{\partial\mathcal{M}} du d\phi \Big[\frac{1}{2}\Phi^{\prime}\dot{\zeta}-\frac{1}{2}(\Phi^{\prime})^{2}+\frac{2\sigma}{3}\eta^{\prime}\dot{\xi}-\frac{2\sigma}{3}(\eta^{\prime})^{2}\Big]
\end{align}

Equation \eqref{eq:final-2Dtheory} is the main result of this paper. This is the action that describes the dynamics of a spin 3 field coupled to three dimensional pure gravity theory. Here $\eta, \xi$ carries the signatures of the higher dimensional higher spin fields and $\sigma$ is the higher spin coupling. To the best of our knowledge, this is the first ever attempt in the literature to write an action describing the dynamics of asymptotically flat higher spin gravity theory in three space time dimensions. We easily note that the canonical conjugate momentas of the above fields can be written as,
\begin{align}
p_{\Phi}=\frac{k}{2}\zeta^{\prime},~ p_{\zeta}=\frac{k}{2}\Phi^{\prime},~p_{\eta}=k\frac{2\sigma}{3}\xi^{\prime},~p_{\xi}=k\frac{2\sigma}{3}\eta^{\prime}.
\end{align}

These canonical conjugate momenta of the fields are used to find the symmetry transformation of the fields. A Hamiltonian formalism can further be used to find the variation of the different fields and that can be computed from their Poisson brackets with the global charge. These variations of fields will help us to get the transformations of various fields that further lead to the symmetries of the reduced action of the theory. This gives us the opportunity to cross-check the asymptotic algebra \eqref{chargealgebra} directly. While the prescription is straightforward, the actual computation requires solving the equations of motions for each field and is technically challenging. It would be nice to address this question in alternative ways and we hope to report on it in a future work. 

\section{Discussion and Outlook}\label{sec6}

The asymptotic symmetry algebra of asymptotically flat 3D spacetimes, $\bmst$ admits generic deformations known as $W(a,b)$ algebra. In this paper, we have established a no-go theorem, that for a generic value of the parameter $b$ other than -1, $W(0,b)$ algebras can not be used as a gauge group for a three dimensional Chern Simons theory, since they do not admit a non-degenerate invariant bilinears. Next, we studied the possible emergence of $W(a,b)$ algebras as the asymptotic symmetry algebra (or at least a subalgebra of an asymptotic algebra) of a  gravity theories in 3D flat spacetimes. In particular, a key observation of our work is that $W(0,-2)$ algebra can be thought of as a \emph{subalgebra} of the asymptotic symmetry algebra of asymptotically flat spin-3 gravity. Specifically, the generators $L_{\alpha}$ and $V_{p}$ form the $W(0,-2)$ subalgebra. Having thus identified a connection between deformations of $\bmst$ and symmetry algebra of asymptotically flat higher spin theories for a particular case (of $a=0$ and $b=-2$), it is natural to ask if generalization of such connection exists for generic values of $a, b$ and arbitrary higher spin. We hope to report on this issue in future works. 

As a natural subsequent step, we considered spin-3 gravity theory in the bulk and use the Chern-Simons formulation of (2+1)D gravity to study its asymptotic symmetries with suitable boundary conditions imposed on all fields. Alternatively, we also derive the algebra as an {I}n\"{o}n\"{u}-Wigner contraction of the asymptotic symmetry algebra for spin-3 gravity in AdS$_3$. In a sense the {I}n\"{o}n\"{u}-Wigner contraction is tantamount to taking the large radius limit of $AdS$. Our results of the charge algebra are in agreement with the existing one of the literature \cite{Afshar:2013vka}. Through this computation, the appearance of symmetric combinations of boundary fields in the gauge variations becomes evident and finds its root in the corresponding {I}n\"{o}n\"{u}-Wigner contraction prescription.

Another central result of this paper is to obtain the explicit action of the corresponding 2D dual theory for this bulk higher spin extended gravity theory. The prime tool used in deriving the 2D theory is the equivalence between the Chern-Simons theory and Wess-Zumino-Witten (WZW) theory. In deriving the explicit form of the action we have used three ingredients: (I) Polyakov-Wiegmann identity which gives us a decomposition of the Chern-Simons action as the sum of three independent Chern-Simons theories along with quartic and sextic interaction terms. (II) Gaussian decomposition of gauge group elements played a crucial role in the derivation of the simplified 2D action. It is in fact this decomposition that motivates the particular Polyakov-Wiegmann decomposition performed in the previous section. (III) Finally, the boundary conditions on the chemical potentials in the Chern-Simons theory is used, which goes into defining the independent degrees of freedoms of the dual two dimensional theory. Eventually, the \emph{chiral} WZW theory, is reduced to a flat limit of an extended Liouville theory, where the fields $\eta, \xi$ carry the signatures of the higher dimensional higher spin fields. This gives rise to a relatively simple looking 2D theory \eqref{eq:final-2Dtheory}, that describes the dynamics of the three dimensional asymptotically flat gravity theory coupled to spin 3 fields. A careful symmetry analysis of this theory is bounded to reproduce the asymptotically flat spin-3 algebra although we relegate that analysis to future works. It is noteworthy to mention that one might attempt to compute the exact (resummed) $S$-matrix element of the higher spin Liouville theory obtained in this work using integrability techniques developed recently in \cite{Jorjadze:2020uow,Jorjadze:2021ily}.

\paragraph{Acknowledgement :}
SB would like to thank the organizers of the “Quantum Black Holes, Quantum Information and Quantum Strings” workshop held at Asia Pacific Center for Theoretical Physics (APCTP), Pohang, South Korea for giving an opportunity to present this work and for the fruitful discussions on the same. The work of AM is supported by the Ministry of Education, Science, and Technology (NRF- 2021R1A2C1006453) of the National Research Foundation of Korea (NRF). The work of DM is supported by the Department of Science and Technology, Govt. of India grant number SB/SJF/2019-20/08. Finally, we would like to thank the people of India for their generous support towards research in basic sciences.

\appendix
\section{Polyakov-Wiegmann identity}

In this appendix, we present the Polyakov-Wiegmann identity in a covariant form. Our starting point is the following action:
\begin{equation}
\label{eq:WZW}
    I(g)=\frac{k}{2}\int_{\Sigma} d^2 \sigma\ \sqrt{-\rho}\rho^{ij}\ \text{Tr}(g^{-1}\partial_i g\ g^{-1}\partial_j g)+\frac{k}{3}\int_V d^3 \sigma\ \varepsilon^{ijk}\text{Tr}(G^{-1}\partial_i G\ G^{-1}\partial_j G\ G^{-1}\partial_k G)\ ,
\end{equation}
where $g$ is a group element of the gauge group associated with the above action, $\Sigma$ is the 2-dimensional manifold, $V$ is the 3-dimensional bulk extension of $\Sigma$ (i.e. $\partial V= \Sigma$) and $G$ is the extension of $g$ in the 3-dimensional bulk. We also assume that the coordinates of the 2-dimensional manifold $\Sigma$ are both compact. Consider the following decomposition of $g$, namely 
\begin{equation}
\label{eq:gdecomposition}
    g=f\cdot h\ .
\end{equation}
Correspondingly, let $F$ and $H$ be the extensions of $f$ and $h$ in the 3-dimensional bulk manifold $V$. Plugging \eqref{eq:gdecomposition} in the first term of \eqref{eq:WZW},
\begin{equation}
    \text{Tr}(g^{-1}\partial_i g\ g^{-1}\partial_j g)=\text{Tr}(f^{-1}\partial_i f\ f^{-1}\partial_j f)+\text{Tr}(h^{-1}\partial_i h\ h^{-1}\partial_j h)+2\text{Tr}(f^{-1}\partial_i f\ \partial_j h h^{-1})\ .
\end{equation}
In writing the above, we have also used the cyclicity of traces. Now, plugging in 
\begin{equation}
    G=F\cdot H
\end{equation}
in the bulk term of \eqref{eq:WZW}, we can again use cyclicity of traces to get
\begin{equation}
\begin{aligned}
    \varepsilon^{ijk}\text{Tr}(G^{-1}\partial_i G\ G^{-1}\partial_j G\ G^{-1}\partial_k G)&=\varepsilon^{ijk}\text{Tr}(F^{-1}\partial_i F\ F^{-1}\partial_j F\ F^{-1}\partial_k F)\\
    &+\varepsilon^{ijk}\text{Tr}(H^{-1}\partial_i H\ H^{-1}\partial_j H\ H^{-1}\partial_k H)\\
    &+3\varepsilon^{ijk}\text{Tr}(\partial_i H\ H^{-1} F^{-1}\partial_j F\ F^{-1}\partial_k F)\\
    &+3\varepsilon^{ijk}\text{Tr}(\partial_i H\ H^{-1} \partial_j H\ H^{-1} F^{-1}\partial_k F)
\end{aligned}
\end{equation}
Focussing on the last two terms, note that 
\begin{equation}
    \varepsilon^{ijk}\text{Tr}(\partial_i H\ H^{-1} F^{-1}\partial_j F\ F^{-1}\partial_k F+\partial_i H\ H^{-1} \partial_j H\ H^{-1} F^{-1}\partial_k F)=-\partial_j(\varepsilon^{ijk}\partial_i H\ H^{-1}\ F^{-1}\partial_k F)\ ,
\end{equation}
implying that they are precisely boundary pieces. Thus, eventually, we have
\begin{equation}
\begin{aligned}
    I(g)&=I(f)+I(h)+k\int_{\Sigma} d^2 \sigma\ \sqrt{-\rho}\rho^{ij}\text{Tr}(f^{-1}\partial_i f \partial_j h h^{-1})\\
    &\hspace*{6cm}-k\int_V d^3\sigma\ \text{Tr}\left( \partial_j (\varepsilon^{ijk}F^{-1}\partial_k F\ \partial_i H H^{-1})\right)\\
    &=I(f)+I(h)+k\int_{\Sigma} d^2 \sigma\ \sqrt{-\rho}\rho^{ij}\text{Tr}(f^{-1}\partial_i f \partial_j h h^{-1})\\
    &\hspace*{6cm}+k\int_\Sigma d^2 \sigma\ \sqrt{-\rho}\varepsilon^{ik}\text{Tr}(f^{-1}\partial_k f\ \partial_i h h^{-1})\label{de1}
\end{aligned}
\end{equation}

The components of $\rho$  in light-cone coordinate are $\rho^{+-}=\rho^{-+}=2$ and $\rho^{++}=\rho^{--}=0$. Similarly for $\varepsilon$, the components are $\varepsilon^{+-}=2$, $\varepsilon^{-+}=-2$ and $\varepsilon^{++}=\varepsilon^{--}=0$. $\rho$ and $\varepsilon$ are the determinats of the $\rho_{ij}$ and $\varepsilon_{ij}$.\\
\begin{equation}
\begin{aligned}
    I(g)
    &=I(f)+I(h)+2k\int_{\Sigma} d^2 x \text{Tr}(f^{-1}\partial_{-} f \partial_{+} h h^{-1})
\end{aligned}
\end{equation}

Now, further decomposing $f$ in \eqref{de1} as
\begin{equation}
    f=p \cdot q \ ,
\end{equation}
and their bulk extensions as
\begin{equation}
    F=P \cdot Q \ ,
\end{equation}
we get,
\begin{equation}
\begin{aligned}
    I(g)=&I(p)+I(q)+I(h)+k\int_{\Sigma} d^2 \sigma\ \sqrt{-\rho}\rho^{ij}\text{Tr}(p^{-1}\partial_i p\ \partial_j q q^{-1})\\
    &+k\int_V d^3\sigma\ \text{Tr}\left( \partial_j (\varepsilon^{ijk}P^{-1}\partial_k P\ \partial_i Q Q^{-1})\right)\\
    &+k\int_{\Sigma} d^2 \sigma\ \sqrt{-\rho}\rho^{ij}\text{Tr}(q^{-1}p^{-1}\partial_i (pq) \partial_j h h^{-1})\\
    &+k\int_V d^3\sigma\ \text{Tr}\left( \partial_j (\varepsilon^{ijk}Q^{-1}P^{-1}\partial_k (PQ)\ \partial_i H H^{-1})\right) 
\end{aligned}
\end{equation}
\begin{equation}
\begin{aligned}
    \Rightarrow I(g)=&I(p)+I(q)+I(h)+k\int_{\Sigma} d^2 \sigma\ \sqrt{-\rho}\rho^{ij}\text{Tr}(p^{-1}\partial_i p\ \partial_j q q^{-1})\\
    &+k\int_{\Sigma} d^2 \sigma\ \sqrt{-\rho}\rho^{ij}\text{Tr}(q^{-1}\partial_i q\ \partial_j h h^{-1})+k\int_{\Sigma} d^2 \sigma\ \sqrt{-\rho}\rho^{ij}\text{Tr}(p^{-1}\partial_i p\ q\partial_j h h^{-1} q^{-1})\\
    &+k\int_{\Sigma} d^2 \sigma \sqrt{-\rho}\varepsilon^{ij}\text{Tr}(p^{-1}\partial_j p\ \partial_i q q^{-1})+k\int_{\Sigma} d^2 \sigma \sqrt{-\rho}\varepsilon^{ij}\text{Tr}(q^{-1}\partial_j q\ \partial_i h h^{-1})\\
    &+k\int_{\Sigma} d^2 \sigma \sqrt{-\rho}\varepsilon^{ij}\text{Tr}(p^{-1}\partial_j p\ q\partial_i h h^{-1} q^{-1})
\end{aligned}
\end{equation}
\begin{equation}
\begin{aligned}
    \Rightarrow I(g)=&I(p)+I(q)+I(h)+2k\int_{\Sigma} d^2 x \text{Tr}(p^{-1}\partial_{-} p\ \partial_{+} q q^{-1})+2k\int_{\Sigma} d^2 x \text{Tr}(q^{-1}\partial_{-} q\ \partial_{+} h h^{-1})\\
    &+2k\int_{\Sigma} d^2 x\ \text{Tr}(p^{-1}\partial_{-} p\ q\partial_{+} h h^{-1} q^{-1})
\end{aligned}
\end{equation}

The final form of Polyakov-Wiegmann identity is given as,
\begin{align}
    I(g)=&I(p)+I(q)+I(h)+2k\int_{\Sigma} du d\phi \text{Tr}(p^{-1}\partial_{\phi} p\ \partial_{u} q q^{-1})+2k\int_{\Sigma} du d\phi \text{Tr}(q^{-1}\partial_{\phi} q\ \partial_{u} h h^{-1}) \nonumber\\
    &+2k\int_{\Sigma} du d\phi\ \text{Tr}(p^{-1}\partial_{\phi} p\ q\partial_{u} h h^{-1} q^{-1})
\end{align}

We have used the above form of the identity for the computations of the dual gravity action. 

\section{Dual Action for pure gravity}
In this appendix, we have re-derived the dual Wess-Zumino-Witten action for three dimensional pure asymptotically flat  Einstein gravity theory. We have used the Polyakov-Wiegmann identity for the derivation. The Wess-Zumino-Witten action which is equivalent to Chern-Simons action in $(u,\phi,r)$ coordinate can be given as,
\begin{align}
    I_{CS}[g]=k\int_{\partial \mathcal{M}} du d\phi \langle \partial_{\phi} g g^{-1} \partial_{u}g g^{-1}\rangle + \frac{k}{3}\int_{\mathcal{M}} \langle G^{-1}dG \wedge G^{-1}dG \wedge G^{-1}dG \rangle + I_{bdy}
\end{align}

We can decompose element $g$ into  three parts i.e. $g=g_{1}.g_{2}.g_{3}$ and using the Polyakov-Wiegmann identity. We can write the action as,
\begin{align}  I_{CS}(g)&=I(g_{1})+I(g_{2})+I(g_{3})+2k\int_{\partial\mathcal{M}} du d\phi \text{Tr}(g_{1}^{-1}\partial_{\phi} g_{1}\ \partial_{u} g_{2} g_{2}^{-1}) \nonumber\\
    &+2k\int_{\partial\mathcal{M}} du d\phi \text{Tr}(g_{2}^{-1}\partial_{\phi} g_{2}\ \partial_{u} g_{3} g_{3}^{-1})+2k\int_{\partial\mathcal{M}} du d\phi\ \text{Tr}(g_{1}^{-1}\partial_{\phi} g_{1}\ g_{2}\partial_{u} g_{3} g_{3}^{-1} g_{2}^{-1})+I_{bdy}
\end{align}

For $\bmst$ case, $g_{1},g_{2},g_{3}$ can be expressed as an exponential sum of the generators of the Poincare algebra (iso(2,1)) with a field coefficient can be given as,
\begin{align}\label{gdp}
    g_{1}&=e^{X L_{1}+W M_{1}}\nonumber\\
     g_{2}&=e^{\Phi L_{0}+\zeta M_{0}}\nonumber\\
     g_{3}&=e^{Y L_{-1}+V M_{-1}}
\end{align}

where $\{L_{i},M_{i}\}(i=0,\pm 1)$ are the generators of the iso(2,1) algebra and $X, W, \Phi, \zeta, Y, V$ is the corresponding fields. The iso(2,1) algebra can be given as,
    \begin{align}
    [L_{m},L_{n}]=&(m-n)L_{m+n} \nonumber \\
    [L_{m},M_{n}]=&(m-n)M_{m+n}
\end{align}
 explicit matrix representation of this iso(2,1) algebra in three dimensions can be written as follows,
 \begin{align}
 L_{1}&=
     \begin{pmatrix}
         0 & 0 & 0 \\
         1 & 0 & 0 \\
         0 & 1 & 0
     \end{pmatrix}, 
      L_{-1}=
     \begin{pmatrix}
         0 & -2 & 0 \\
         0 & 0 & -2 \\
         0 & 0 & 0
     \end{pmatrix}, 
      L_{0}=
     \begin{pmatrix}
         1 & 0 & 0 \\
         0 & 0 & 0 \\
         0 & 0 & -1
     \end{pmatrix}
 \end{align}
 and we have defined $M_{i}= \epsilon L_{i}(i=0,\pm 1)$ where $\epsilon^{2}=0$. The non-degenarate supertrace element of the generators of iso(2,1) is given as, $\langle L_{m},M_{n} \rangle =~$anti-diag$(-\frac{1}{2},\frac{1}{4},-\frac{1}{2})$.

 Using the above relation, we can show that,
 \begin{align}
     I(g_{1})&=0 \\
     I(g_{2})&=k \int_{\mathcal{M}} du d\phi~\Big(\frac{1}{4}\Phi^{\prime}\dot{\zeta}+\frac{1}{4}\zeta^{\prime}\dot{\Phi} \Big)\\
     I(g_{3})&=0
 \end{align}
 and 
 \begin{align}
      &\text{Tr}(g_{1}^{-1}\partial_{\phi} g_{1}\ \partial_{u} g_{2} g_{2}^{-1})=0\\
      &\text{Tr}(g_{2}^{-1}\partial_{\phi} g_{2}\ \partial_{u} g_{3} g_{3}^{-1})=0\\
      &\text{Tr}(g_{1}^{-1}\partial_{\phi} g_{1}\ g_{2}\partial_{u} g_{3} g_{3}^{-1} g_{2}^{-1})= -\frac{1}{2} e^{\Phi} X^{\prime} \dot{V}-\frac{1}{2} e^{\Phi} W^{\prime} \dot{Y}-\frac{1}{2} e^{\Phi} \zeta X^{\prime} \dot{Y}
 \end{align}

 So the Action can be written as,
 \begin{align}
     I_{CS}=& k \int_{\mathcal{M}} du d\phi \Big[ \frac{1}{4}\Phi^{\prime}\dot{\zeta}+\frac{1}{4}\zeta^{\prime}\dot{\Phi}+ 2\Big(-\frac{1}{2} e^{\Phi} X^{\prime} \dot{V}-\frac{1}{2} e^{\Phi} W^{\prime} \dot{Y}-\frac{1}{2} e^{\Phi} \zeta X^{\prime} \dot{Y}\Big)\Big]+ I_{bdy}\nonumber\\
     =& -\frac{k}{2} \int_{\mathcal{M}} du d\phi \Big[ 2\Big(e^{\Phi} X^{\prime} \dot{V}+e^{\Phi} W^{\prime} \dot{Y}+e^{\Phi} \zeta X^{\prime} \dot{Y}\Big) - \frac{1}{2}(\Phi^{\prime}\dot{\zeta}+\zeta^{\prime}\dot{\Phi})\Big]+ I_{bdy}
 \end{align}

This result exactly matches with the result of \cite{Merbis:2019wgk} up to an overall factor.
\section{Matrix representation of \texorpdfstring{$\mathfrak{fhs}(3)$}{} algebra }
\color{black}
An explicit $3\times3$ matrix representation of the generators, that we have used in this paper is chosen as follows,
 \begin{align}
 L_{1}&=
     \begin{pmatrix}
         0 & 0 & 0 \\
         1 & 0 & 0 \\
         0 & 1 & 0
     \end{pmatrix}, 
      L_{-1}=
     \begin{pmatrix}
         0 & -2 & 0 \\
         0 & 0 & -2 \\
         0 & 0 & 0
     \end{pmatrix}, 
      L_{0}=
     \begin{pmatrix}
         1 & 0 & 0 \\
         0 & 0 & 0 \\
         0 & 0 & -1
     \end{pmatrix}, \nonumber\\
     U_{0}&=\frac{2}{3}\sqrt{-\sigma}
      \begin{pmatrix}
         1 & 0 & 0 \\
         0 & -2 & 0 \\
         0 & 0 & 1
     \end{pmatrix}, 
      U_{1}=\sqrt{-\sigma}
     \begin{pmatrix}
         0 & 0 & 0 \\
         1 & 0 & 0 \\
         0 & -1 & 0
     \end{pmatrix}, 
      U_{-1}=\sqrt{-\sigma}
     \begin{pmatrix}
         0 & -2 & 0 \\
         0 & 0 & 2 \\
         0 & 0 & 0
     \end{pmatrix},\nonumber\\
     U_{2}&=2\sqrt{-\sigma}
     \begin{pmatrix}
         0 & 0 & 0 \\
         0 & 0 & 0 \\
         1 & 0 & 0
     \end{pmatrix}, 
      U_{-2}=2\sqrt{-\sigma}
     \begin{pmatrix}
         0 & 0 & 4 \\
         0 & 0 & 0 \\
         0 & 0 & 0
     \end{pmatrix},
 \end{align}
 and defined $M_{i}= \epsilon L_{i}(i=0,\pm 1)$ and $V_{i}= \epsilon U_{i}(i=0,\pm 1,\pm 2)$ where $\epsilon^{2}=0$. $\langle L_{m},M_{n} \rangle =$anti-diag$(-\frac{1}{2},\frac{1}{4},-\frac{1}{2})$ and $\langle U_{m},V_{n} \rangle =$anti-diag$(2\sigma,-\frac{\sigma}{2},\frac{\sigma}{3},-\frac{\sigma}{2},2\sigma)$, where $\sigma$ is a constant.

\bibliographystyle{hieeetr}
\bibliography{References}

\end{document}

%% file: gdefn.tex
\usepackage{epsfig,graphicx,bm,amsmath,amsfonts}
\usepackage{epstopdf}
\usepackage{amssymb,xcolor,float}
\usepackage{amsthm}
\usepackage{hyperref}
\usepackage{cite}
\usepackage{float}
\usepackage{longtable}
\usepackage{parskip}
\usepackage{graphicx,epsfig}
\usepackage{psfrag}
\usepackage{wrapfig}

\usepackage{xcolor}
\usepackage[english]{babel}
\usepackage[T1]{fontenc}
\usepackage[utf8]{inputenc}
\usepackage{authblk}
\usepackage{mathtools}
\usepackage{slashed}
\usepackage{mattens,amssymb}
\usepackage{amsmath,amssymb}
\usepackage{amsfonts}
\usepackage{graphicx,color}
\usepackage{cite}
\usepackage{hyperref}
\hypersetup{
	colorlinks=true,
	linkcolor=blue,
	filecolor=magenta,      
	urlcolor=cyan,
}
\usepackage[capitalize]{cleveref}
\numberwithin{equation}{section} 

\definecolor{refcol}{rgb}{0.9,0.1,0.1}
\hypersetup{colorlinks=true,linkcolor=blue,citecolor=refcol,urlcolor=cyan,linktocpage}

\usepackage{wrapfig}

\newcommand{\bmst}{\mathfrak{bms}_3}

\newcommand{\ben}{\begin{eqnarray}\displaystyle}
\newcommand{\een}{\end{eqnarray}}

\newcommand{\be}{\begin{equation}}
\newcommand{\ee}{\end{equation}}

\newcommand{\bc}{\begin{center}}
\newcommand{\ec}{\end{center}}

\newcommand{\eesp}{\end{split}}
\newcommand{\bsp}{\begin{split}}

\newcommand{\Rmnum}[1]{\expandafter\@slowromancap\romannumeral #1@}

\newcommand{\cC}{\mathcal{C}}

\newcommand{\cG}{\mathcal{G}}

\newcommand{\bensp}{\begin{eqnarray}\begin{split}}
\newcommand{\eensp}{\end{eqnarray}\end{split}}

\newcommand{\bnm}{\begin{matrix}}
\newcommand{\enm}{\end{matrix}}

\def\XXint#1#2#3{{\setbox0=\hbox{$#1{#2#3}{\int}$ }
\vcenter{\hbox{$#2#3$ }}\kern-.6\wd0}}







%% file: WAlg_HigherSpin.bbl
\begin{thebibliography}{10}

\bibitem{BMvdB:1962}
V.~d. B. M. G.~J. Bondi~Hermann and M.~A.~W. K., ``{1962 Gravitational waves in
  general relativity, VII. Waves from axi-symmetric isolated system},'' {\em
  Proc. R. Soc. Lond. A26921–52}, 1962.

\bibitem{Sachs:1962}
R.~K. Sachs, ``{1962 Gravitational waves in general relativity VIII. Waves in
  asymptotically flat space-time},'' {\em Proc. R. Soc. Lond. A270103–126},
  1962.

\bibitem{Barnich:2011mi}
G.~Barnich and C.~Troessaert, ``{BMS charge algebra},'' {\em JHEP}, vol.~12,
  p.~105, 2011, 1106.0213.

\bibitem{Strominger:2013jfa}
A.~Strominger, ``{On BMS Invariance of Gravitational Scattering},'' {\em JHEP},
  vol.~07, p.~152, 2014, 1312.2229.

\bibitem{He:2014laa}
T.~He, V.~Lysov, P.~Mitra, and A.~Strominger, ``{BMS supertranslations and
  Weinberg\textquoteright{}s soft graviton theorem},'' {\em JHEP}, vol.~05,
  p.~151, 2015, 1401.7026.

\bibitem{Saha:2019tub}
A.~P. Saha, B.~Sahoo, and A.~Sen, ``{Proof of the classical soft graviton
  theorem in $D$ = 4},'' {\em JHEP}, vol.~06, p.~153, 2020, 1912.06413.

\bibitem{Fernandes:2020tsq}
K.~Fernandes and A.~Mitra, ``{Soft factors from classical scattering on the
  Reissner-Nordstr\"om spacetime},'' {\em Phys. Rev. D}, vol.~102, no.~10,
  p.~105015, 2020, 2005.03613.

\bibitem{Pasterski:2017kqt}
S.~Pasterski and S.-H. Shao, ``{Conformal basis for flat space amplitudes},''
  {\em Phys. Rev. D}, vol.~96, no.~6, p.~065022, 2017, 1705.01027.

\bibitem{Fotopoulos:2019vac}
A.~Fotopoulos, S.~Stieberger, T.~R. Taylor, and B.~Zhu, ``{Extended BMS Algebra
  of Celestial CFT},'' {\em JHEP}, vol.~03, p.~130, 2020, 1912.10973.

\bibitem{Banerjee:2021uxe}
N.~Banerjee, T.~Rahnuma, and R.~K. Singh, ``{Asymptotic symmetry of four
  dimensional Einstein-Yang-Mills and Einstein-Maxwell theory},'' {\em JHEP},
  vol.~01, p.~033, 2022, 2110.15657.

\bibitem{Witten:1988hc}
E.~Witten, ``{(2+1)-Dimensional Gravity as an Exactly Soluble System},'' {\em
  Nucl. Phys. B}, vol.~311, p.~46, 1988.

\bibitem{Witten:1988hf}
E.~Witten, ``{Quantum Field Theory and the Jones Polynomial},'' {\em Commun.
  Math. Phys.}, vol.~121, pp.~351--399, 1989.

\bibitem{Moore:1989yh}
G.~W. Moore and N.~Seiberg, ``{Taming the Conformal Zoo},'' {\em Phys. Lett.
  B}, vol.~220, pp.~422--430, 1989.

\bibitem{Elitzur:1989nr}
S.~Elitzur, G.~W. Moore, A.~Schwimmer, and N.~Seiberg, ``{Remarks on the
  Canonical Quantization of the Chern-Simons-Witten Theory},'' {\em Nucl. Phys.
  B}, vol.~326, pp.~108--134, 1989.

\bibitem{Barnich:2013yka}
G.~Barnich and H.~A. Gonzalez, ``{Dual dynamics of three dimensional
  asymptotically flat Einstein gravity at null infinity},'' {\em JHEP},
  vol.~05, p.~016, 2013, 1303.1075.

\bibitem{Banerjee:2015kcx}
N.~Banerjee, D.~P. Jatkar, S.~Mukhi, and T.~Neogi, ``{Free-field realisations
  of the BMS$_{3}$ algebra and its extensions},'' {\em JHEP}, vol.~06, p.~024,
  2016, 1512.06240.

\bibitem{Barnich:2015sca}
G.~Barnich, L.~Donnay, J.~Matulich, and R.~Troncoso, ``{Super-BMS$_{3}$
  invariant boundary theory from three-dimensional flat supergravity},'' {\em
  JHEP}, vol.~01, p.~029, 2017, 1510.08824.

\bibitem{Banerjee:2016nio}
N.~Banerjee, D.~P. Jatkar, I.~Lodato, S.~Mukhi, and T.~Neogi, ``{Extended
  Supersymmetric BMS$_3$ algebras and Their Free Field Realisations},'' {\em
  JHEP}, vol.~11, p.~059, 2016, 1609.09210.

\bibitem{Banerjee:2018hbl}
N.~Banerjee, A.~Bhattacharjee, I.~Lodato, and T.~Neogi, ``{Maximally $
  \mathcal{N} $ -extended super-BMS$_{3}$ algebras and generalized 3D gravity
  solutions},'' {\em JHEP}, vol.~01, p.~115, 2019, 1807.06768.

\bibitem{Banerjee:2019lrv}
N.~Banerjee, A.~Bhattacharjee, Neetu, and T.~Neogi, ``{New $ \mathcal{N} $ = 2
  SuperBMS$_{3}$ algebra and invariant dual theory for 3D supergravity},'' {\em
  JHEP}, vol.~11, p.~122, 2019, 1905.10239.

\bibitem{Banerjee:2019epe}
N.~Banerjee, S.~Khandelwal, and P.~Shah, ``{Equivalent dual theories for 3D
  $\mathcal N=2$ supergravity},'' {\em Phys. Rev. D}, vol.~100, no.~10,
  p.~105013, 2019, 1907.05866.

\bibitem{Banerjee:2021uxl}
N.~Banerjee, A.~Bhattacharjee, S.~Biswas, and T.~Neogi, ``{Dual theory for
  maximally $ \mathcal{N} $ extended flat supergravity},'' {\em JHEP}, vol.~05,
  p.~179, 2022, 2110.05919.

\bibitem{Parsa:2018kys}
A.~Farahmand~Parsa, H.~Safari, and M.~Sheikh-Jabbari, ``{On Rigidity of 3d
  Asymptotic Symmetry Algebras},'' {\em JHEP}, vol.~03, p.~143, 2019,
  1809.08209.

\bibitem{Barnich:2006av}
G.~Barnich and G.~Compere, ``{Classical central extension for asymptotic
  symmetries at null infinity in three spacetime dimensions},'' {\em Class.
  Quant. Grav.}, vol.~24, pp.~F15--F23, 2007, gr-qc/0610130.

\bibitem{Detournay:2012pc}
S.~Detournay, T.~Hartman, and D.~M. Hofman, ``{Warped Conformal Field
  Theory},'' {\em Phys. Rev. D}, vol.~86, p.~124018, 2012, 1210.0539.

\bibitem{Afshar:2019axx}
H.~Afshar, H.~A. Gonz\'alez, D.~Grumiller, and D.~Vassilevich, ``{Flat space
  holography and the complex Sachdev-Ye-Kitaev model},'' {\em Phys. Rev. D},
  vol.~101, no.~8, p.~086024, 2020, 1911.05739.

\bibitem{Compere:2013bya}
G.~Comp\`ere, W.~Song, and A.~Strominger, ``{New Boundary Conditions for
  $\text{AdS}_3$},'' {\em JHEP}, vol.~05, p.~152, 2013, 1303.2662.

\bibitem{Afshar:2015wjm}
H.~Afshar, S.~Detournay, D.~Grumiller, and B.~Oblak, ``{Near-Horizon Geometry
  and Warped Conformal Symmetry},'' {\em JHEP}, vol.~03, p.~187, 2016,
  1512.08233.

\bibitem{Chen:2019hbj}
B.~Chen, P.-X. Hao, and Z.-F. Yu, ``{2d Galilean Field Theories with
  Anisotropic Scaling},'' {\em Phys. Rev. D}, vol.~101, no.~6, p.~066029, 2020,
  1906.03102.

\bibitem{10.1063/1.532067}
J.~Negro, M.~A. del Olmo, and A.~Rodr\'{i}guez-Marco, ``{Nonrelativistic
  conformal groups},'' {\em Journal of Mathematical Physics}, vol.~38,
  pp.~3786--3809, 07 1997,
  https://pubs.aip.org/aip/jmp/article-pdf/38/7/3786/11142288/3786\_1\_online.pdf.

\bibitem{Vasiliev:1990en}
M.~A. Vasiliev, ``{Consistent equation for interacting gauge fields of all
  spins in (3+1)-dimensions},'' {\em Phys. Lett. B}, vol.~243, pp.~378--382,
  1990.

\bibitem{Giombi:2009wh}
S.~Giombi and X.~Yin, ``{Higher Spin Gauge Theory and Holography: The
  Three-Point Functions},'' {\em JHEP}, vol.~09, p.~115, 2010, 0912.3462.

\bibitem{Maldacena:2011jn}
J.~Maldacena and A.~Zhiboedov, ``{Constraining Conformal Field Theories with A
  Higher Spin Symmetry},'' {\em J. Phys. A}, vol.~46, p.~214011, 2013,
  1112.1016.

\bibitem{Campoleoni:2010zq}
A.~Campoleoni, S.~Fredenhagen, S.~Pfenninger, and S.~Theisen, ``{Asymptotic
  symmetries of three-dimensional gravity coupled to higher-spin fields},''
  {\em JHEP}, vol.~11, p.~007, 2010, 1008.4744.

\bibitem{Vasiliev:2012vf}
M.~A. Vasiliev, ``{Holography, Unfolding and Higher-Spin Theory},'' {\em J.
  Phys. A}, vol.~46, p.~214013, 2013, 1203.5554.

\bibitem{Bekaert:2022poo}
X.~Bekaert, N.~Boulanger, A.~Campoleoni, M.~Chiodaroli, D.~Francia,
  M.~Grigoriev, E.~Sezgin, and E.~Skvortsov, ``{Snowmass White Paper: Higher
  Spin Gravity and Higher Spin Symmetry},'' 5 2022, 2205.01567.

\bibitem{Aragone:1983sz}
C.~Aragone and S.~Deser, ``{Hypersymmetry in $D=3$ of Coupled Gravity Massless
  Spin 5/2 System},'' {\em Class. Quant. Grav.}, vol.~1, p.~L9, 1984.

\bibitem{Campoleoni:2021blr}
A.~Campoleoni and S.~Pekar, ``{Carrollian and Galilean conformal higher-spin
  algebras in any dimensions},'' {\em JHEP}, vol.~02, p.~150, 2022, 2110.07794.

\bibitem{Afshar:2013vka}
H.~Afshar, A.~Bagchi, R.~Fareghbal, D.~Grumiller, and J.~Rosseel, ``{Spin-3
  Gravity in Three-Dimensional Flat Space},'' {\em Phys. Rev. Lett.}, vol.~111,
  no.~12, p.~121603, 2013, 1307.4768.

\bibitem{Safari:2019zmc}
H.~R. Safari and M.~M. Sheikh-Jabbari, ``{BMS$_{4}$ algebra, its stability and
  deformations},'' {\em JHEP}, vol.~04, p.~068, 2019, 1902.03260.

\bibitem{Banerjee:2022abf}
N.~Banerjee, A.~Mitra, D.~Mukherjee, and H.~R. Safari, ``{Supersymmetrization
  of deformed BMS algebras},'' 1 2022, 2201.09853.

\bibitem{Polyakov:1984et}
A.~M. Polyakov and P.~B. Wiegmann, ``{Goldstone Fields in Two-Dimensions with
  Multivalued Actions},'' {\em Phys. Lett. B}, vol.~141, pp.~223--228, 1984.

\bibitem{Tsutsui:1994pp}
I.~Tsutsui and L.~Feher, ``{Global aspects of the WZNW reduction to Toda
  theories},'' {\em Prog. Theor. Phys. Suppl.}, vol.~118, pp.~173--190, 1995,
  hep-th/9408065.

\bibitem{Jorjadze:2020uow}
G.~Jorjadze and S.~Theisen, ``{On the S-matrix of Liouville theory},'' {\em
  JHEP}, vol.~02, p.~111, 2021, 2011.06876.

\bibitem{Jorjadze:2021ily}
G.~Jorjadze and S.~Theisen, ``{Generating Functional for the S-Matrix in
  Liouville Theory},'' {\em PoS}, vol.~Regio2020, p.~013, 2021.

\bibitem{Merbis:2019wgk}
W.~Merbis and M.~Riegler, ``{Geometric actions and flat space holography},''
  {\em JHEP}, vol.~02, p.~125, 2020, 1912.08207.

\end{thebibliography}
